\documentclass[longauth]{aa}
\usepackage[colorlinks,citecolor=blue,filecolor=blue,linkcolor=blue,breaklinks=true, plainpages=false,urlcolor=blue]{hyperref}
\usepackage{graphicx,xcolor}
\usepackage[varg]{txfonts}
\usepackage{natbib}
\bibpunct{(}{)}{;}{a}{}{,} 
\usepackage[figuresright]{rotating}

\def\teff{$T\rm_{eff}$ }
\def\kms{$\mathrm{km\, s^{-1}}$}

\newcommand{\mygi}{MyGIsFOS}
\newcommand{\Teff}{\ensuremath{T_\mathrm{eff}}}

\newcommand{\loggf}{\ensuremath{\log\,gf}}
\newcommand{\logg}{\ensuremath{\log g}}

\newcommand{\draftflag}{false}

\newcommand{\beq}{\begin{equation}}
\newcommand{\eeq}{\end{equation}}

\newcommand{\as}{A({\rm S})}

\begin{document}

\title{The Gaia-ESO Survey: Galactic evolution of sulphur and zinc
\thanks{Based on observations collected at the European Organisation for Astronomical Research in the Southern Hemisphere
under ESO programmes 188.B-3002, 193.B-0936.}
}

\author{
S.~Duffau \inst{1,2,3} \and
E.~Caffau \inst{4} \and
L.~Sbordone \inst{5,1,2} \and
P.~Bonifacio \inst{4} \and
S.~Andrievsky \inst{6,4} \and
S.~Korotin \inst{6,7} \and
C.~Babusiaux \inst{4} \and
S.~Salvadori \inst{4} \and
L.~Monaco \inst{3} \and
P.~Fran\c cois \inst{4,8} \and
\'A.~Sk\'ulad\'ottir \inst{9,10} \and
A.~Bragaglia \inst{11} \and
P.~Donati \inst{11,12} \and
L.~Spina \inst{13} \and
A.~J.~Gallagher \inst{4} \and
H.-G.~Ludwig \inst{14,4} \and
N.~Christlieb \inst{14} \and
C.~J.~Hansen \inst{15,16} \and 
A.~Mott \inst{17} \and
M.~Steffen \inst{17} \and
S.~Zaggia \inst{18} \and
S.~Blanco-Cuaresma \inst{19} \and
F.~Calura \inst{11} \and
E.~Friel \inst{20} \and
F.~M.~Jim\'enez-Esteban \inst{21} \and
A.~Koch \inst{22} \and
L.~Magrini \inst{23} \and
E.~Pancino \inst{23,24} \and
B.~Tang \inst{25} \and
G.~Tautvai\v{s}ien\.{e} \inst{26} \and
A.~Vallenari \inst{18} \and
K.~Hawkins \inst{27}
G.~Gilmore\inst{28}\and
S.~Randich \inst{23} \and
S.~Feltzing \inst{29} \and
T. ~Bensby \inst{29} 
E.~Flaccomio \inst{34} \and
R.~Smiljanic \inst{30} \and
A.~Bayo \inst{31} \and
G.~Carraro \inst{32} \and
A.R.~Casey \inst{28} \and
M.~T.~Costado \inst{33} \and
F.~Damiani \inst{34} \and
E.~Franciosini \inst{23} \and
A.~Hourihane \inst{28} \and
P.~Jofr\'e \inst{28,35} \and
C.~Lardo \inst{36} \and
J.~Lewis \inst{28} \and 
L.~Morbidelli \inst{23} \and
S.~G.~Sousa \inst{37} \and
C.~C.~Worley \inst{28} 
}

\institute{Millennium Institute of Astrophysics, Santiago, Chile
\and
Instituto de Astrof\'isica, Pontificia Universidad Cat\'olica de Chile, Av. Vicu\~na Mackenna 4860, 782-0436 Macul, Santiago, Chile
\and 
Departamento de Ciencias Fisicas, Universidad Andres Bello, Fernandez Concha 700, Las Condes, Santiago, Chile
\and
GEPI, Observatoire de Paris, PSL Research University, CNRS, Place Jules Janssen, 92195 Meudon, France
\and
European Southern Observatory, Alonso de Cordova 3107, Vitacura, Santiago de Chile, Chile
\and
Department of Astronomy and Astronomical Observatory, Odessa National University, Isaac Newton Institute of Chile, Odessa Branch, Shevchenko Park, 65014, Odessa, Ukraine
\and
Crimean Astrophysical Observatory, Nauchny 298409, Republic of Crimea
\and
UPJV, Universit\'e de Picardie Jules Verne, 33 Rue St Leu, F-80080 Amiens
\and
Kapteyn Astronomical Institute, University of Groningen, Landleven 12, NL-9747 AD Groningen, The Netherlands
\and
Max-Planck-Institut f\" ur Astronomie, K\" onigstuhl 17, D-69117 Heidelberg, Germany
\and
INAF - Osservatorio Astronomico di Bologna, via Gobetti 93/3, 40129 Bologna, Italy
\and
Dipartimento di Astronomia, Universit\'a di Bologna, via Gobetti 93/2, 40129 Bologna, Italy
\and
Universidade de Sao Paulo, IAG, Departamento de Astronomia, Rua do Matao 1226, Sao Paulo, 05509-900, SP, Brasil
\and
Zentrum f\"ur Astronomie der Universit\"at Heidelberg, Landessternwarte, K\"onigstuhl 12, 69117 Heidelberg, Germany
\and
Technische Universität Darmstadt, Schlossgartenstr. 2, Darmstadt D-64289, Germany
\and
Dark Cosmology Centre, The Niels Bohr Institute, Juliane Maries Vej 30, 2100 Copenhagen, Denmark e-mail: cjhansen@dark-cosmology.dk
\and
Leibniz-Institut f{\"u}r Astrophysik Potsdam, An der Sternwarte 16, D-14482 Potsdam, Germany
\and
Istituto Nazionale di Astrofisica, Osservatorio Astronomico di Padova Vicolo dell'Osservatorio 5, 35122 Padova, Italy
\and
Observatoire de Gen\`eve, Universit\'e de Gen\`eve, CH-1290 Versoix, Switzerland
\and
Department of Astronomy, Indiana University, 727 East 3rd St Swain West 318 Bloomington,  IN  47405, US
\and
Centro de Astrobiolog\'{\i}a (INTA-CSIC), Departamento de Astrof\'{\i}sica, PO Box 78, E-28691, Villanueva de la Ca\~nada, Madrid, Spain
\and
Phyics Department, Lancaster University, Lancaster LA1 4YB, UK
\and
INAF - Osservatorio Astrofisico di Arcetri, Largo E. Fermi 5, 50125, Florence, Italy
\and
ASI Science Data Center, Via del Politecnico SNC, 00133 Roma, Italy
\and
Departamento de Astronom\'ia, Casilla 160-C, Universidad de Concepci\'on, Chile
\and
Institute of Theoretical Physics and Astronomy, Vilnius University, Sauletekio av. 3, 10222 Vilnius, Lithuania
\and
Department of Astronomy, Columbia University, 550 W 120th St, New York, NY 10027
\and
Institute of Astronomy, University of Cambridge, Madingley Road, Cambridge CB3 0HA, United Kingdom
\and
Lund Observatory, Department of Astronomy and Theoretical Physics, Box 43, SE-221 00 Lund, Sweden
\and
Nicolaus Copernicus Astronomical Center, Polish Academy of Sciences, ul. Bartycka 18, 00-716, Warsaw, Poland
\and
Instituto de F\'isica y Astronom\'ia, Universidad de Valpara\'iso, Chile
\and
Dipartimento di Fisica e Astronomia, Universit\'a di Padova, Vicolo Osservatorio 3, I-35122, Padova, Italy
\and
Instituto de Astrof\'{i}sica de Andaluc\'{i}a-CSIC, Apdo. 3004, 18080 Granada, Spain
\and
INAF - Osservatorio Astronomico di Palermo, Piazza del Parlamento 1, 90134, Palermo, Italy
\and
N\'ucleo de Astronom\'ia, Facultad de Ingenier\'ia, Universidad Diego Portales,  Av. Ejercito 441, Santiago, Chile
\and
Laboratoire d'astrophysique, Ecole Polytechnique F\'ed\'erale de Lausanne (EPFL), Observatoire de Sauverny, CH-1290 Versoix, Switzerland
\and
Instituto de Astrof\'isica e Ci\^encias do Espa\c{c}o, Universidade do Porto, CAUP, Rua das Estrelas, 4150-762 Porto, Portugal
}
\authorrunning{Duffau et al.}
\titlerunning{Gaia-ESO Survey S and Zn}
\date{Received ...; Accepted ...}

\abstract
%
{Due to their volatile nature, when sulphur and zinc are observed in external galaxies, their determined abundances represent the gas-phase abundances
in the interstellar medium. 
 This implies that they can be used as tracers of the chemical enrichment of matter in the Universe at high redshift.
Comparable observations in stars are more difficult and, until recently, plagued by
small number statistics.}
{We wish to  exploit the Gaia ESO Survey (GES) data  to study 
the behaviour of sulphur and zinc 
abundances of a large number of Galactic stars, in a homogeneous way.}
{By using the  UVES spectra of the GES sample, 
we are able to assemble a sample of 1301 Galactic stars, including stars in open and globular clusters
in which both sulphur and zinc were 
measured.}
{We confirm the results from the literature that sulphur behaves as an $\alpha$-element.
We find a large scatter in [Zn/Fe] ratios among giant stars around solar metallicity. 
The lower ratios are observed in giant stars at Galactocentric distances less than 7.5\,kpc.
No such effect is observed among dwarf stars, since they do not extend to that radius. 
}
{
Given the sample selection, giants and dwarfs are observed at
different Galactic locations, and it is plausible, and compatible with simple calculations, 
that Zn-poor giants trace a younger population more polluted by SN Ia yields.
It is necessary to extend observations in order
to observe both giants and dwarfs at the same Galactic location. Further theoretical
work on the evolution of zinc is also necessary.}
\maketitle


\section{Introduction}


It is known from observations of the Galaxy's interstellar medium (ISM) that many chemical elements in the gas phase can be depleted into 
dust grains (such elements are also refereed to as refractory). Zinc and sulphur are
two of the few elements that are relatively unaffected by this in the ISM \citep[i.e. volatile, see e.g.][]{savage96}.
This makes sulphur and zinc interesting elements to investigate, and for this reason
we make use of the large sample of stars observed by the Gaia-ESO Survey 
\citep[GES,][]{gesmessenger,randich13}, for which both sulphur and zinc have been analysed,
to investigate the abundances of these relatively volatile elements in Galactic stars and to exploit their potential as tracers of Galactic chemical evolution.


\subsection{Sulphur}
Sulphur is produced in the final stage of the evolution of massive stars,
type II supernovae \citep[SNe,][typically on timescales of less than 30 Myr]{woosley95,limongi03,chieffi04}.
On the other hand, the elements of the iron-peak are produced by type II SNe but mainly 
in type Ia SNe (SN Ia), which produce little to no $\alpha$-elements 
\citep{nomoto84,iwamoto99}.
To picture the evolution of the stellar populations in a galaxy it is important to derive 
the chemical abundances of both the $\alpha$-elements and iron-peak elements. In fact, abundance ratios 
between elements formed on different timescales, such as [$\alpha$/Fe], can be used as cosmic
clocks and allow us to clarify the star formation history of our targets.
The first phases of evolution of a galaxy are characterised by a low content of metals, 
because only a small number of massive stars have had enough time to evolve
and explode and/or transfer mass to a companion which can in turn reach the mass limit and
explode. In both cases, they enrich the environment 
with the metals synthesised during their stellar life.
This early environment is mainly characterised by enrichment from type II SNe
but there is hardly any contribution from type Ia SNe, 
possibly with little traces of the products of the promptest type\,Ia explosions at evolutionary times greater than 30-40 Myr \citep[see][]{manucci06,greggio83}.
The metal-poor environment is then characterised by an over abundance of $\alpha$-elements
with respect to iron-peak elements when compared to the Sun. 
This is usually referred to as the $\alpha$-elements
enhancement, generally observed in metal-poor stars \citep[e.g.][]{venn04,cayrel04,bonifacio09,hayden15}
and also theoretically predicted \citep{tinsley79,matteucci90}.
At the solar metallicity regime, on the other hand, the ratio of $\alpha$-elements over iron is the same as in the Sun,
because type Ia SNe have
had time to explode and enrich the ISM with iron-peak elements.

A few \ion{S}{i} features are detected as absorption lines in the spectra of late-type stars.
Sulphur lines typically used to derive the sulphur abundance belong to five 
\ion{S}{i} multiplets (Mult.)\footnote{We adopt the multiplet numbering of \citet{Moore}.}.
Mult.\,1 at 920\,nm, Mult.\,3 at 1045\,nm, Mult.\,6 at 869\,nm, Mult.\,8 at 670\,nm and Mult.\,10 at 605\,nm.
A forbidden line at 1082\,nm can be used only to analyse giant stars. This line has an equivalent width
of about 0.2\,pm in the Solar spectrum \citep{zolfito} therefore it can be used to derive 
$\as$\footnote{$A{\rm (X)}=\log_{10}{\left( \frac{ N{\rm \left(X\right)} }{ N{\rm \left(H\right)}} \right)} + 12$.} 
in dwarf stars only at metal-rich regime in extremely high quality spectra. 
Two decades after the pioneering investigation of sulphur by \citet{GeorgeW},
who analysed nine stars in six globular clusters,
the systematic analysis of sulphur established itself \citep{fran87,fran88}.

Sulphur, as an $\alpha$-element, is expected to scale with iron in solar metallicity stars 
(${\rm [S/Fe]}$\footnote{$\left[{\rm X/Y}\right]=\log_{10}{\left(\frac{N({\rm X})}{N({\rm Y})}\right)}_* - \log_{10}{\left(\frac{N({\rm X})}{N({\rm Y})}\right)}_\sun$}$\approx 0.0$), to increase with respect to iron (${\rm [S/Fe]}> 0.0$) as 
metallicity decreases in the range of $-1.0<${\rm [Fe/H]}$<-0.3$, and to have a quite constant 
positive value of about ${\rm [S/Fe]}\approx +0.4$ in the metal-poor regime ${\rm [Fe/H]}<-1.0$. 
In the theoretical framework, \citet{chiappini99} and \citet{kobayashi06} investigated the Galactic evolution 
of sulphur by calculating its evolution in the solar neighbourhood, adopting their own nucleosynthesis
yields. They obtained their yields based on the new developments at that time 
in the observational and theoretical studies of SNe and
extremely metal-poor stars in the Galactic halo. In Fig.\,11 of \citet{kobayashi06}, 
an almost flat [S/Fe] is predicted in the range $-3.0<{\rm [Fe/H]}<-1.0$.
Unlike other $\alpha$-elements, like silicon and calcium, sulphur is relatively volatile. 
This in turn means that it is not locked 
into dust grains that form in the ISM \citep[but see][]{calura09,jenkins09}.
As a consequence the sulphur abundance derived in stellar photospheres can be
directly compared to the sulphur present in the ISM,
for instance derived from emission lines in spectra of Blue Compact Galaxies
and from resonance absorption lines in Damped Ly-$\alpha$ systfems
\citep{garnett89,centurion00}.

Although all the recent works agree on an average increase of [S/Fe] below solar metallicity,
not all studies agreed on a constant [S/Fe] at metal-poor regime. Instead, various scenarios have been reported: 
{a constant increase of [S/Fe] as metallicity decreases \citep[see][]{israelian01,takadahidai02}};
an increase followed by a flat [S/Fe] at the metal-poor regime as metallicity decreases \citep{nissen04,nissen07}; 
 a bimodal behaviour of [S/Fe] at the metal-poor regime \citep{zolfo05}.
Most recent papers tend to agree on a flat [S/Fe] in metal-poor stars \citep{spite11,matrozis13}.

After the pioneering work of \citet{caffau05b}, who investigated sulphur in three stars belonging 
to Terzan\,7 -- a globular cluster belonging to the Sagittarius dwarf Spheroidal galaxy -- 
sulphur abundance has been derived in several open and globular clusters.
Recently \citet{skuladottir15} analysed the \ion{S}{i} lines of Mult.\,1 in a sample of
85 stars in the Sculptor dwarf spheroidal galaxy, in the metallicity range $-2.5<{\rm [Fe/H]}<-0.8$.
They found that sulphur behaves as the other
$\alpha$-elements in Sculptor, when effects due to departures from local thermodynamical equilibrium (NLTE) are taken into account.

In Tables\,\ref{comp-tab} and \,\ref{comp-tab2} we summarise some important literature results 
for abundance determination of sulphur in the Galaxy for field stars, stars in open and globular clusters, 
and stars in some extra-galactic objects as well\footnote{In listing [Fe/H] and [S/Fe] in these tables we did {\em not} attempt to
homogenize the respective solar abundance scales, since all the values listed are indicative and would not vary in any significant way}.
The metallicity ranges, as well as the observed trends (flat, sloping etc.), are indicated and comments regarding the data have 
been added for clarity.

Sulphur has also been derived in the Apache Point Observatory Galactic Evolution Experiment (APOGEE) project.
APOGEE is part of the third phase of the Sloan Digital sky survey (SDSS-III). 
A near-infrared spectrograph ($1.51-1.70\mu$m; $R=22\,500$) was used to collect spectra for about
$150\,000$ stars over three years \citep[][]{holtzman15}. Stellar
atmospheric parameters were released, together with abundance data for 15
elements, including sulphur \citep[see Fig. 14 in][]{holtzman15} derived from the \ion{S}{i} lines at 1.5\,$\mu$m.
\citet[][]{hayden15} selected about $70\,000$ cool giants from SDSS data release
12 and studied the behaviour of the Milky Way disk in the ${\rm [\alpha/Fe]}$ versus
[Fe/H] plane over a large volume, namely within 2\,kpc from the plane, at
radial distance $3\leq{\rm R}\leq15$\,kpc of the galactic centre (bulge excluded). 
Together with O, Mg, Si, Ca and Ti, sulphur is one of the elements which are combined
to provide the [$\alpha$/Fe] parameter in APOGEE. \citet[][]{hayden15} found
that in the inner disk (${\rm R}<5$\,kpc), stars are distributed along a single sequence,
while in the outer disk (${\rm R}>5$\,kpc), stars distributed along a high and a low
$\alpha$ sequence (see their figures 3 and 4). 
Such a double sequence was predicted theoretically by \citet{caluramenci09}.
\citet{hayden15} found the shape of the high
$\alpha$ sequence to remain constant with the radial distance, although very few
of them were found at ${\rm R}>11$\,kpc \citep[see also][]{nidever14}.
The high and low $\alpha$ sequences are indicated to correspond to the thick and
thin disk populations \citep[][]{holtzman15}.

Some special stars of relevance have also been investigated. 
\citet{bonifacio12}, analysing the Mult.\,3, could derive only an upper 
limit for the sulphur content in HE\,1327-2326 \citep[one of the most iron-poor stars known][]{frebel05}. 
\citet{roederer16} derived the first abundance of sulphur in a
carbon-enhanced metal-poor star {\citep[below {[Fe/H]}=$-3$, but see also][]{skuladottir15}} by analysing BD+44$^\circ$493.
At this iron abundance regime (${\rm [Fe/H]}=-3.8$), the usually investigated \ion{S}{i} features are too weak.
They investigated three ultra-violet \ion{S}{i} lines at 181\,nm
in a spectrum observed with HST using the Cosmic Origins Spectrograph.
They derived [S/Fe]$=+0.07\pm 0.41$. 
This value is compatible with the behaviour of [S/Fe] in metal-poor stars,
albeit with large uncertainty.

Two independent NLTE investigations 
can be found in \citet{takeda05} and \citet{korotin08,korotin09}. They conclude that 
NLTE effects are usually large for the strong lines of Mult.\,1 and 3,
while they are much smaller for the weak lines of Mult.\,6 and 8.

\begin{table*}
 \centering
 \scriptsize
\caption{Summary of literature regarding Galactic sulphur abundances}
\label{comp-tab}
\begin{tabular}{@{}l@{}cc@{}c@{}ccccc@{}}
\hline\hline
Reference & Mult. & N$_{obs}$ & NLTE & target & Flat & Slope & Comments \\
 & line &   &  & type & range & range &  \\
\hline
\citet{clegg81} &  6  & 20 &  - & F\&G MS &                  -                  & $-0.9 < \rm{ [Fe/H] } < +0.4$    &  large scatter \\ 
\citet{fran87}   &  6  & 13 &  - &  MP dwarfs  & $\rm {[Fe/H] } < -0.5$  &  $-0.5 < \rm{ [Fe/H] } < +0.3$ &  1 halo star\\ 
\citet{fran88}   &  6  & 12 &  - &  field dwarfs        & $-1.3 < \rm{ [Fe/H] } < -0.5$ &                   -                    &  5 are halo stars\\
\citet{takadahidai96} &  6  & 11 & $\surd$ &  peculiar stars & solar &                   -                    &  \\
\citet{israelian01} & 6 & 8 & - &        MP         &                  -                    & $-3.0 < \rm{ [Fe/H] } < -0.6$ & negligible NLTE \\ 
\citet{takadahidai02} &  6  & 68 & $\surd$ & field stars & - & {\bf $-3 < \rm{ [Fe/H] } < 0.0$}   &  \\
\citet{chen02} & 6,8,10 & 26 & - & disc stars & $\rm -1.0 < {[Fe/H] } < -0.5$  & $-0.5 < \rm{ [Fe/H] } < +0.5$ & - \\ 
\citet{chen03} & 6,8,10 & 15 & - & old MR & - &  $-0.1 < \rm{ [Fe/H] } < +0.5$ & - \\ 
\citet{ecuvillon04} & 8 & 112/31 & - & planet/no planet & - & $-0.8 < \rm{ [Fe/H] } < +0.5$ & - \\ 
\citet{nissen04} & 1,8 & 34 & - & dwarf/subgiant halo & $-3.2 < \rm{ [Fe/H] } < -0.8$ & - & 3D effects small, two deviant stars \\ 
\citet{ryde04}  & 1 & 10 & - & - & $-3.0 < \rm{ [Fe/H] } < -0.7$ & - & - \\ 
\citet{korn05} & 1,6 & 3 & - & MP & $-2.43 < \rm{ [Fe/H] } < -2.08$ & - & - \\
\citet{zolfo05} & 1,6,8 & 74 & - & - & $-3.2 < \rm{ [Fe/H] } < -1.0$ (*) & $-1.0 < \rm{ [Fe/H] } < +0.5$ & in combined lit.sample (253 stars) \\ 
                       &          &      &    &    &                                                                                              &  & (*) dual behaviour at [Fe/H] < $-$1 \\
\citet{takeda05} & 1,3,6 & 3 & $\surd$ & FGK Stars & $-3.17 < \rm{ [Fe/H] } < -2.0$ (*) & $-2.0 < \rm{ [Fe/H] } < +0.47$ & in combined lit. sample (175 stars) \\
                       &          &      &    &    &                                                                                              &  & (*) dual behaviour at [Fe/H] < $-$2  \\
\citet{ryde06} & $\rm[SI]$  & 14 & - & disk giants/subgiants & - & $-0.66 < \rm{ [Fe/H] } < +0.03$ & - \\
\citet{nissen07} & 1,6 & 40 & $\surd$ & MS halo & $-3.3 < \rm{ [Fe/H] } < -1.0$ & - & low scatter \\ 
                         & 3     & 1  & - & - & $\rm{ [Fe/H] } = -1.69$ & - & - \\ 
\citet{ecmult3} & 3 & 5 & $\surd$ & F-K MS & - & - & test for Mult.3 analysis,good \\
                        &    &    &              &              &    &   & agreement with Mult. 6 and 8 results \\
\citet{caffau10} & 3 & 4 & $\surd$ & MP halo & - & - & no flat behaviour, scatter \\
                         &     &    &             &               &    &    & in $-2.42 < \rm{ [Fe/H] } < -1.19$\\
\citet{takeda11}  & 3 & 33 & $\surd$ & halo/disk & $-2.5 < \rm{ [Fe/H] } < -1.5$ & $-1.0 < \rm{ [Fe/H] } < 0.0$ & jump in $\as$ with [Fe/H] \\
                           &      &    &              &                &                                              &              &  $-3.7 < \rm{ [Fe/H] } < -2.5$\\
\citet{takeda12}  & 3 & 13 & $\surd$ & TO dwarfs \& giants &  $-3.2 < \rm{ [Fe/H] } < -1.9$ & - & - \\
\citet{spite11} & 1 & 33 & $\surd$ & EMP & $-3.5 < \rm{ [Fe/H] } < -2.9$ & - & - \\
\citet{jonsson11} & 3, $\rm[SI]$ & 10 & $\surd$, - & giants & $-3.4 < \rm{ [Fe/H] } < -1.5$ & - & $\as$ from [SI] larger than from Mult. 3  \\
\citet{matrozis13} & $\rm [SI]$ & 39 & - & MP giants & $-2.3 < \rm{ [Fe/H] } < -1.0$ & $-1.0 < \rm{ [Fe/H] } < 0.0$ & - \\
\citet{takeda16} & 6,8 & 239 & $\surd$ & giants & - & $-0.8<{\rm [Fe/H]}<+0.2$ & high scatter (Mult. 6), $\as$ from Mult. 8 \\
                          & 6,8 & 160 & $\surd$ & dwarfs & -&  $-1.3<{\rm [Fe/H]}<+0.5$ & $\as$ from Mult. 8 \\
\citet{giano16}  & 3 & 4  & $\surd$ & dwarfs & - & - & - \\
\hline
\end{tabular}
\end{table*}

\begin{table*}
 \centering
 \scriptsize
\caption{Summary of literature regarding sulphur abundances in open clusters, globular clusters and extra-galactic stars.}
\label{comp-tab2}
\begin{tabular}{@{}l@{}ccccccc@{}}
\hline\hline
Reference & Mult. & N$_{obs}$ &  Object & $\rm {[Fe/H]} $ & $\rm {[S/Fe]} $ & Consistent & Comments \\
 & line &   &  & dex & dex & with MW? & \\
\hline
\citet{caffau05b}    & 1 & 3 & Terzan 7 (Sgr dSph)   & $-0.5$ & $-0.05$ & - & Lower than MW, \\
                              &     &    &                                    &        &          &     &   consistent with Sgr dSph \\
\citet{sbordone09} & 1 & 4 & NGC 6752   & $-1.43$ & ${\rm \langle [S/Fe] \rangle }= +0.49 \pm 0.15$ & $\surd$ & - \\
                              & 1 & 9 & 47 Tuc         & $-0.67$ & ${\rm \langle [S/Fe] \rangle }= +0.18 \pm 0.14$ & $\surd$ & S-Na corr. \\
\citet{koch11}         &  1 & 1 & NGC 6397   & $-2.1$ & ${\rm \langle [S/Fe] \rangle }= +0.52 \pm 0.20$ & $\surd$ & - \\
\citet{caffau14}       &  1 & 10 & M4        & $-1.36$ & 0.51 & $\surd$ & - \\
                       &  1 & 1 & Trumpler 5 & $-0.53$ &  $-0.23$ & - & low $\rm {[S/Fe]}$ with NLTE \\
                       &  8 & 4 & NGC 5822   & $0.0$ & ${\rm \langle [S/Fe] \rangle } \approx -0.02$ & $\surd$ & - \\
                       &  8 & 7 & NGC 2477   & $0.0$ & ${\rm \langle [S/Fe] \rangle } \approx -0.04$ & $\surd$ & - \\
\citet{kacharov15}   &   3 & 6 & M4            &  $-1.08$ & ${\rm \langle [S/Fe] \rangle }= +0.58 \pm 0.20$ & $\surd$ & no S-Na corr. \\
                                &   3 & 6 & M22         & $-1.9<{\rm [Fe/H]}<-1.6$ & ${\rm \langle [S/Fe] \rangle }= +0.57 \pm 0.19$ & $\surd$ & 1 star high $\rm {[S/Fe]}$, no S-Na corr. \\
                                &  3 & 3 & M30          & $-2.3$ & ${\rm \langle [S/Fe] \rangle }= +0.55 \pm 0.16$ & $\surd$ & 1 stars high $\rm {[S/Fe]}$, no S-Na corr. \\
\citet{skuladottir15} &  1 & 85 & Sculptor dSph & $-2.5<{\rm [Fe/H]}<-2.0$ & $\rm {[S/Fe]} \approx$ +0.16 & $\surd$ & NLTE included \\
                               &   &  &  & $-2.0<{\rm [Fe/H]}<-0.8$ & $\rm {[S/Fe]} $ decreasing w/ increasing $\rm {[Fe/H]}$  & $\surd$ & NLTE included \\
\hline
\end{tabular}
\end{table*}

\subsection{Zinc}
Zinc, also a relatively volatile element, was believed to be an accurate tracer of 
iron-peak elements, since ${\rm [Zn/Fe]}\approx0$ in early measurements of thin disk stars 
(e.g. \citealt{snedencrocker88,sneden91}). Zinc has thus been historically used to infer the 
iron content of the gas in Damped Lyman-$\alpha$ systems 
\citep[DLAs, e.g. see][for a complete review -- in particular their Sec. 3.2]{wolfe05}.
In principle, measurements of [S/Zn] versus [Zn/H] in stellar photospheres can thus be 
directly compared with the abundances of distant DLAs, which might represent the ISM of formation of these stellar
populations \citep[e.g. see][for a recent comparison]{berg15}.

However, [Zn/Fe] is not constant at different [Fe/H] for stars in the thin and thick disks 
\citep{prochaska00,reddy03,reddy06,bensby03,bensby05,bensby13,bensby14,allendeprieto04}, 
the Galactic halo \citep{primas00,bihain,cayrel04,nissen04,nissen07,bonifacio09},
the bulge \citep{barbuy15} and nearby dwarf galaxies
\citep{sbordone07,venn12,skuladottir16}. This implies that zinc is not necessarily a good
tracer of iron and that its nucleosynthetic origin is probably more
complex than previously thought.

More specifically, the observed [Zn/Fe] is super-solar in the Milky 
Way halo, reaching ${\rm [Zn/Fe]} \approx +0.5$ at ${\rm [Fe/H]} < -3$ \citep{cayrel04,nissen07}. 
In the thin and thick disks, 
[Zn/Fe] decreases with increasing metallicity, reaching the solar ratio at 
$\rm [Fe/H] \approx 0$ similar to the $\alpha$-elements 
\citep{reddy03,reddy06,bensby03,bensby05,bensby13,bensby14}. 
Furthermore, \citet{nissenschuster11} identified two populations in the solar 
neighbourhood with low- and high-$\alpha$ abundances, which also showed low and high ratios 
of [Zn/Fe], respectively. In the dwarf galaxies Sagittarius, Carina and Sculptor, 
measurements of [Zn/Fe] have revealed sub-solar ratios and possible scatter 
\citep{sbordone07,venn12,skuladottir16}.
While there are only very sparse measurements of Zn in ultra-faint dSphs to date, the
few data above [Fe/H]$>-2.2$ that overlap with our sampled range
agree with our Galactic stars. Likewise, the remainder of the very metal-poor ultra-faint
dSph stars with [Fe/H]$<-2$ fully agree with the metal-poor halo 
\citep{frebel10,frebel14,frebel16,roederer16,ji16}.

The fact that [Zn/Fe] increases at low metallicities, albeit
at much lower metallicities than does the [$\alpha$/Fe] ratio,
seems to suggest a related origin of $\alpha$-elements and Zn.
However, predicted zinc yields of SNe Type\,II 
are too low to be compatible with ratios of $\rm [Zn/Fe] \gtrsim 0$ \citep{nomoto97a}. 
By invoking models of core collapse SNe with high explosion energy, i.e. hypernova, which are 
predicted to produce high levels of zinc, \citet{kobayashi06} were able to reproduce the trend 
observed in the disk. This does not, however, explain the high levels of [Zn/Fe] at the lowest 
metallicities observed in the Milky Way halo. To produce the decreasing trend with metallicity, 
the ratio of [Zn/Fe] in SNe Type\,Ia yields should be even lower, and they are indeed predicted 
to be extremely low \citep{iwamoto99}. In addition to hypernova and SNe Type\,II other production 
sites of zinc have been proposed, such as neutrino-driven winds \citep{hoffman96} and the weak 
and/or main $s$-processes, but these are not expected to be dominant sources 
of zinc in the Milky Way \citep{mishenina02,travaglio04}.

Furthermore, there is still no clear consensus on the observed zinc abundances 
at the highest metallicities ($\rm [Fe/H] \gtrsim 0$). Some studies have reported 
an increase in [Zn/Fe] with [Fe/H] at this metallicity \citep{bensby03,bensby05,allendeprieto04}, 
while other suggest a flatter trend \citep{pompeia03,reddy03,reddy06,bensby13,bensby14}. 
\citet{barbuy15} observed red giant stars in the Milky Way bulge, and found a spread of 
$-0.6< \rm [Zn/Fe]<+0.15$ for $\rm [Fe/H]\geq -0.1$, which has not been observed in dwarf stars. 
On the other hand, \citet{takeda16} did not find any significant scatter, nor a discrepancy 
of measured zinc abundances between field dwarf and giant stars in this same metallicity range.

\subsection{The role of the Gaia ESO Survey}
The Gaia ESO Survey provides us with a large, homogeneous sample of
Galactic stars that can help us to understand the Galactic chemical evolution of sulphur and zinc around
solar metallicity. The only features of \ion{S}{i} observable in the VLT-UVES \citep{dekker00} ranges 
observed by GES ($415-621$\,nm or $472$-$683$\,nm, red arm 520\,nm and 580\,nm standard settings, respectively) are weak, and belong to Mult.\,8, which comprises three features, each consisting of a triplet.
Due to the relatively low signal-to-noise ratio ($S/N$) and the relatively
low resolving power ($R\approx47\,000$) of the observed spectra, only the stronger sulphur triplet at 675.7\,nm
is useful for abundance determination in this case.
As stated by \citet{korotin09}, NLTE effects are relatively small for this triplet, making our data very useful for this work. 
Two \ion{Zn}{i} lines are observable in the wavelength range, at 481.0 and 636.2\,nm.

The structure of the paper is as follows: In section 2, we describe the data. 
Section 3 presents the chemical abundance analysis for each element and some specific findings, 
while in Section 4 we discuss these findings and summarise our results.

\section{The Gaia-ESO data}

The Gaia ESO Survey is one of ESO's large public spectroscopic surveys. It is an ambitious project which aims at collecting and analysing high
quality spectra for about 10$^5$ stars by the time the survey will be completed, 
which will complement the spectroscopic capabilities of the Gaia satellite \citet{prusti16} at
faint magnitudes. The GES provides high quality information about the
kinematics and chemistry of the Milky Way bulge, the thin and thick disk, the
halo and of a selected sample of about 60 open clusters covering a range of
ages and masses. Additionally, data is collected for a
number of clusters and benchmark stars for calibration purpose \citep[][]{pancino16}. 
The project is run at the Paranal Observatory on the Chilean
Andes, using the FLAMES \citep{pasquini02} multi-object facility mounted at the UT2 telescope of
the VLT.

The data analysed in the present paper belong to the fourth internal release (henceforth iDR4). 
iDR4 includes observations from the beginning of the survey (31 December 2011)
until end July 2014. In this paper, only the UVES part of the collected
data is considered. UVES observations are conducted mainly with the 580\,nm setup which covers the
wavelength range 472-683\,nm. The fiber target allocation is summarised in \citet[][]{smiljanic14,stonkute16}. 
F- and G-type dwarf stars are the primary targets in solar neighbourhood fields and should cover
distances up to 2\,kpc from the Sun, while a smaller selection of giants extends to larger distances.
In globular clusters, all the targets belong to the RGB or the red clump. In open clusters, red clump stars 
are the main UVES targets in old and intermediate age open clusters, with F-G dwarfs being also observed mainly
in young cluster, and close intermediate-age ones.
The data reduction of the UVES spectra \citep{sacco14} makes use of the ESO
FLAMES-UVES CPL
pipeline\footnote{\url{http://www.eso.org/sci/software/pipelines/}}. 
A detailed description of the structure of the GES UVES sample, and of the strategy adopted to
analyse it, is presented in \citet{smiljanic14}.

\section{Chemical abundance analysis}

The UVES spectra have been analysed using the multiple pipelines strategy described in \citet{smiljanic14}. 
The individual results of the pipelines are combined with an updated methodology to define a final set of 
recommended values of the atmospheric parameters and abundances \citep[see][]{casey16}.
We adopted here the homogenised stellar parameters from GES iDR4. 

With fixed stellar parameters (effective temperature, surface gravity, micro-turbulence
and [Fe/H]) at the values derived in the homogenised iDR4, we ran \mygi\ \citep{sbordone14}
to derive $\as$ from a line profile fitting of \ion{S}{i} Mult.\,8, located at 675\,nm. 
We chose to redetermine S abundances
rather than employ the GES homogenised values due to an extra \ion{S}{i} component mistakenly introduced in the second 
version of the GES line-list, which would skew 
the homogenised results towards lower S abundances.

\begin{figure}
\begin{center}
\resizebox{\hsize}{!}{\includegraphics[draft = \draftflag, clip=true]
{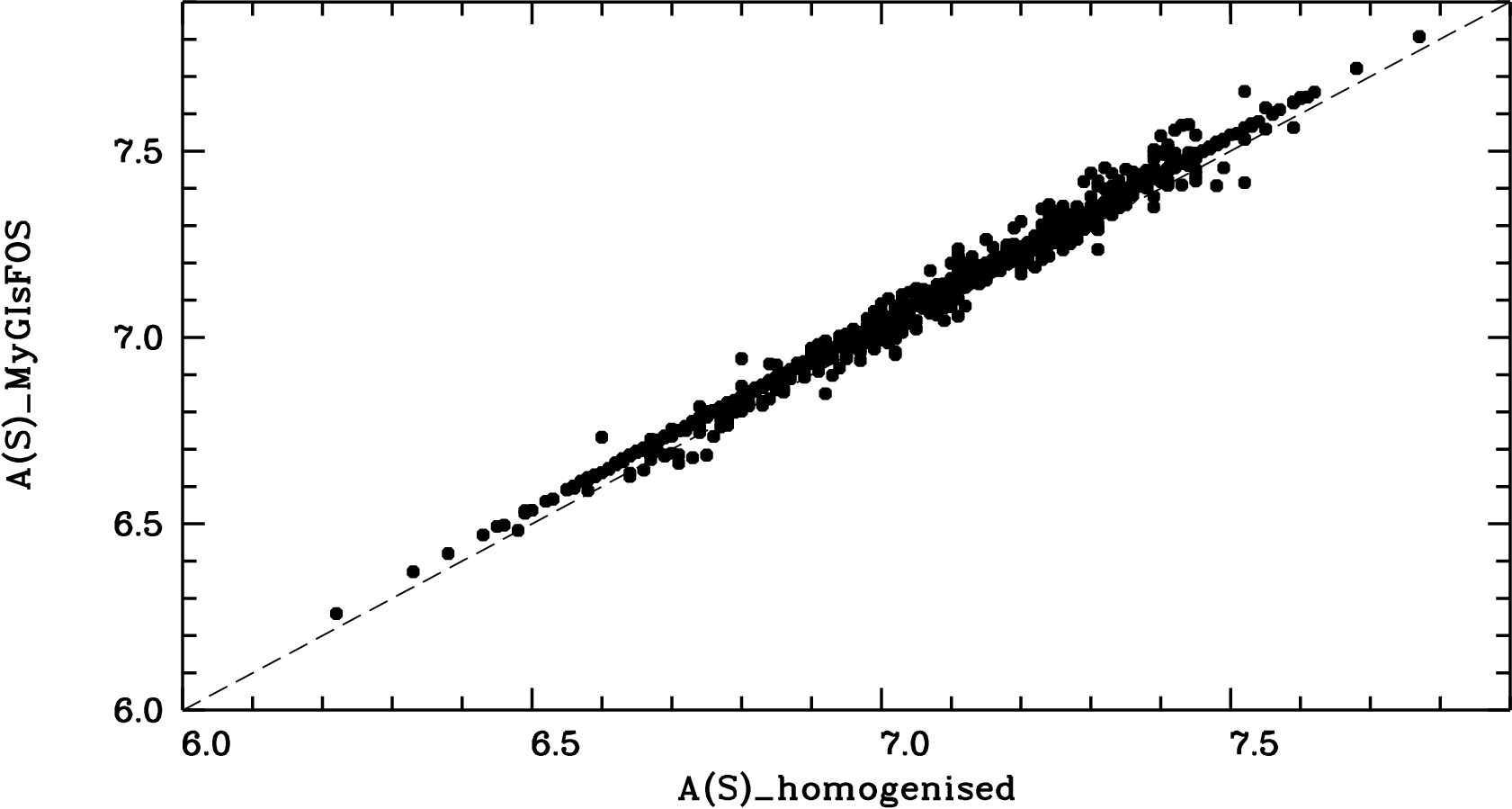}}
\end{center}
\caption[]{Comparison of the $\as$ values found in the present work versus the homogenised values provided by the GES. 
The systematic average difference among the two abundances is  at $-0.03$\,dex, small compared to the typical uncertainties
in the abundance determination.}
\label{homo}
\end{figure}

A grid of synthetic spectra computed with {\sf turbospectrum} \citep{alvarez98,plez12},
based on the grid of OSMARCS models provided by the GES collaboration \citep{smiljanic14}
were fit the observed \ion{S}{i} triplet.
The employed atomic
data are presented in Table\,\ref{irpline}. The \loggf\ of the \ion{S}{i} components used by \citet{takeda16} are slightly larger
than the values chosen by the GES collaboration and 
correspond to about a global \loggf\ 0.01\,dex larger.
The values suggested in NIST provide a global \loggf\ 0.11\,dex lower that would provide
larger sulphur abundances.
A comparison of  the GES homogenised S abundances with the one we derived (giving a systematic difference of $-0.03$\,dex) is presented in Fig.\,\ref{homo}.
{The S abundances we derived are made available as an online table through CDS.}

\begin{table}
\caption{\ion{S}{i} and \ion{Zn}{i} lines analysed in this work.}
\label{irpline}
\begin{center}
\begin{tabular}{lclr}
\hline\hline
\noalign{\smallskip}
 Element & $\lambda$ & E$_{\rm low}$ & \loggf  \\
         & {[nm]}         & {[eV]}\\
\noalign{\smallskip}\hline
\noalign{\smallskip}
\ion{S}{i}  & 675.6750 & 7.87 & $-1.67$ \\
\ion{S}{i}  & 675.6960 & 7.87 & $-0.83$ \\
\ion{S}{i}  & 675.7150 & 7.87 & $-0.24$ \\
\noalign{\smallskip}\hline
\noalign{\smallskip}
\ion{Zn}{i} & 481.0528 & 4.078 & $-0.16$ \\ 
\ion{Zn}{i} & 636.2338 & 5.796 & $+0.14$ \\
\noalign{\smallskip}
\hline
\end{tabular}
\end{center}
\end{table}

\begin{figure}
\begin{center}
\resizebox{\hsize}{!}{\includegraphics[draft = \draftflag, clip=true]
{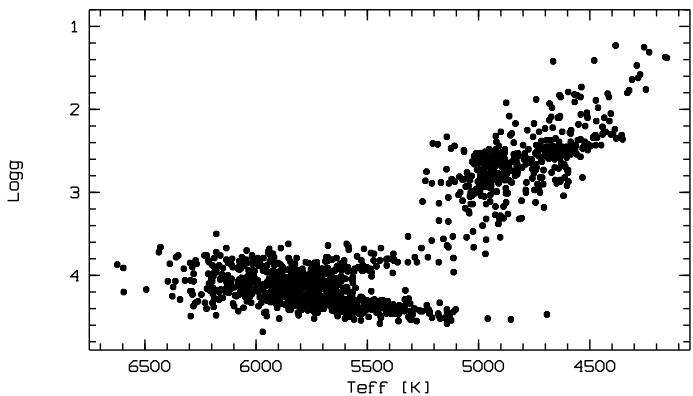}}
\end{center}
\caption[]{The stars analysed in this work are shown in the \teff\ vs. \logg\ plane.}
\label{param}
\end{figure}

We visually inspected all the UVES spectra of F, G and K stars, and retained spectra for which a safe
measurement of the \ion{S}{i} line could be made. We ended up with a sample of 1301 stars.
Our sample spans 2741\,K in effective temperature ($4153\le{T\rm_{eff}}\le 6624$\,K), 3.45\,dex 
in surface gravity ($1.23\le{\logg}\le 4.68$), and 1.68\,dex in metallicity ($-1.07<{\rm [Fe/H]}<+0.61$).
In Fig.\,\ref{param}, the effective temperature and surface gravity of our sample of stars are shown.
{To assure us that this way of selecting the stars did not introduce a bias, we compared [Zn/Fe] in the
complete sample and the selected ones and we could not find any systematic difference (see Sec.\,3.3).}

\begin{figure}
\begin{center}
\resizebox{\hsize}{!}{\includegraphics[draft = \draftflag, clip=true]
{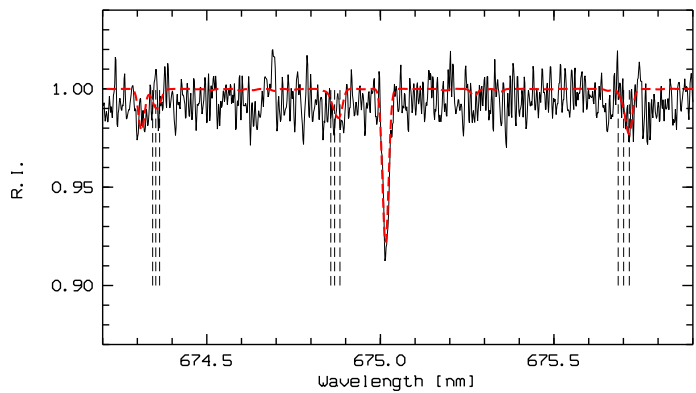}}
\end{center}
\caption[]{The observed spectrum (solid black) of the star 21300738+1210330 
in the range of the Mult.\,8 of sulphur. A synthetic (dashed red) spectrum
with the sulphur abundance derived of $\as=6.26$ by the analysis is also included.
Vertical dashed black lines highlight the positions of the \ion{S}{i} lines of the Mult.\,8.
}
\label{smp}
\end{figure}

One star (21300738+1210330, a member of the cluster M\,15) 
with ${\rm [Fe/H]}=-2.65$, shows a feature at the wavelength of the sulphur triplet (see Fig.~\ref{smp})
but due to a $S/N$ of 100, which is low when compared to the weak line, 
we cannot exclude this as a spurious result. 

We have relied on the homogenised abundances provided by iDR4 for zinc (and also compared to the abundances from our MyGIsFOS node) and the other elements shown here. The abundances are derived from two \ion{Zn}{i} lines at 481.0 and 636.2\,nm.
For the majority of the stars in the sample, zinc abundances are derived from both lines, but for 92 stars the abundances rely on a single
\ion{Zn}{i} line. The zinc lines used are listed in Table\,\ref{irpline}.
The 636.2\,nm line is affected by the \ion{Ca}{i} 636.1\,nm auto-ionisation line and also blending CN lines \citep[see][for a discussion]{barbuy15}.

The solar abundances we have used for reference in our analysis are presented in Table\,\ref{solar_table}.
In the table we also provide the corresponding iDR4 recommended values from GES analysis of an UVES solar spectrum.

\begin{table}
\caption{Solar abundances used in this work compared with GES homogenised values}
\label{solar_table}
\begin{center}
\begin{tabular}{lclr}
\hline\hline
\noalign{\smallskip}
 Element & GES iDR4 & Adopted & reference \\
\noalign{\smallskip}\hline
\noalign{\smallskip}
$A({\rm Fe})$  & 7.43 & 7.52 & \citet{caffau11}  \\
$A({\rm S})$    & 7.03 & 7.16 & \citet{caffau11}  \\
$A({\rm Zn})$  & 4.47 & 4.62 &  \citet{lodders09} \\
$A({\rm Ca})$ & 6.21 & 6.33 &  \citet{lodders09} \\ 
$A({\rm Na})$ & 6.21 & 6.30 &  \citet{lodders09} \\
\noalign{\smallskip}
\hline
\end{tabular}
\end{center}
\end{table}

\subsection{Sulphur}

\begin{figure}
\begin{center}
\resizebox{\hsize}{!}{\includegraphics[draft = \draftflag, clip=true]
{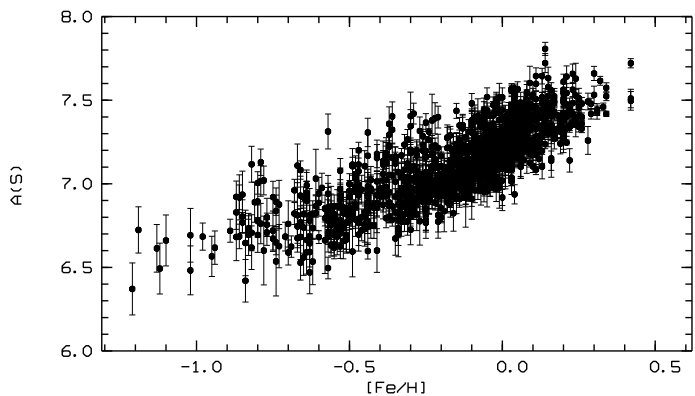}}
\end{center}
\caption[]{$\as$ vs. [Fe/H]. Error bars take into account 3\,$\sigma$ uncertainty
due to spectral $S/N$ and derived by using Cayrel's formula.
}
\label{shplot}
\end{figure}

The spectra we investigate are usually of good quality.
For the spectra for which we derive sulphur abundance, we have a mean signal-to-noise ratio,
in the sample of spectra, of $\langle{S/N}\rangle = 76$ at 580\,nm, with a dispersion 
around the mean value of 19. Only 37 spectra have a $S/N$ lower than 40.
The $S/N$ induces an uncertainty on the sulphur abundance that we quantify
with Cayrel's formula \citep{cayrel88}, and by taking a $3\sigma$ interval
we obtain an average uncertainty of $0.065\pm 0.032$\, dex.
In Fig.\,\ref{shplot} the abundance of sulphur is displayed as a function
of the stellar metallicity and the error bars refer to the uncertainty related to the $S/N$.

The uncertainties in the stellar parameters lead to an uncertainty in the sulphur abundance determination.
We considered five representative stars and derived the impact of changes in \teff\ and \logg\ on
the determination of $\as$. The results are presented in Table\,\ref{errori}.
\begin{table}
\caption{Uncertainties on $\as$ related to uncertainties in the stellar parameters.}
\label{errori}
\begin{center}
\begin{tabular}{@{}lrrrr@{}}
\hline\hline
\noalign{\smallskip}
Stellar parameters & $\Delta\as$ & $\Delta\as$ & $\Delta\as$ & $\Delta\as$ \\
(\teff/\logg/[Fe/H]) & {$\pm 100$\,K} & {$\pm 200$\,K} & {$\pm 0.2$\,dex} & {$\pm 0.4$\,dex} \\
\noalign{\smallskip}\hline
\noalign{\smallskip}
4700/2.5/--0.25 & $^{-0.09}_{+0.10}$& $^{-0.12}_{+0.23}$& $^{+0.06}_{-0.06}$& $^{+0.13}_{-0.11}$\\
\noalign{\smallskip}
4950/2.6/--0.30 & $^{-0.08}_{+0.10}$& $^{-0.16}_{+0.23}$& $^{+0.06}_{-0.07}$& $^{+0.14}_{-0.15}$\\
\noalign{\smallskip}
5300/4.4/--0.50 & $^{-0.09}_{+0.07}$& $^{-0.13}_{+0.16}$& $^{+0.10}_{-0.11}$& $^{+0.24}_{-0.19}$\\
\noalign{\smallskip}
5800/4.3/+0.0   & $^{-0.06}_{+0.06}$& $^{-0.11}_{+0.13}$& $^{+0.07}_{-0.07}$& $^{+0.15}_{-0.13}$\\
\noalign{\smallskip}
6300/4.1/--0.30 & $^{-0.04}_{+0.04}$& $^{-0.07}_{+0.09}$& $^{+0.06}_{-0.06}$& $^{+0.10}_{-0.10}$\\
\noalign{\smallskip}
\hline
\end{tabular}
\end{center}
\end{table}
The uncertainties associated with temperature and gravity in the iDR\,4 sample are $112$\,K and 0.22\,dex, respectively.
Propagation of these errors imply an uncertainty in $\as$ of about $\pm 0.10$\,dex in both cases.

\begin{figure}
\begin{center}
\resizebox{\hsize}{!}{\includegraphics[draft = \draftflag, clip=true]
{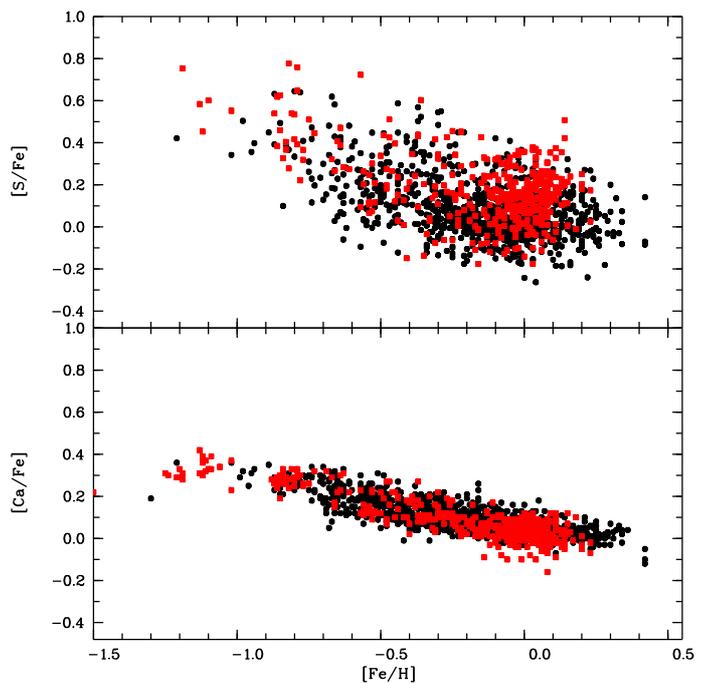}}
\end{center}
\caption[]{[S/Fe] vs. [Fe/H] (upper panel) compared to [Ca/Fe] vs. [Fe/H] (lower panel).
Black filled circles refer to dwarfs while red filled squares to giant stars.
Error bars are not included for clarity, but they are presented in Fig.\,\ref{plotsfe2}.
}
\label{plotsfe}
\end{figure}

\begin{figure}
\begin{center}
\resizebox{\hsize}{!}{\includegraphics[draft = \draftflag, clip=true]
{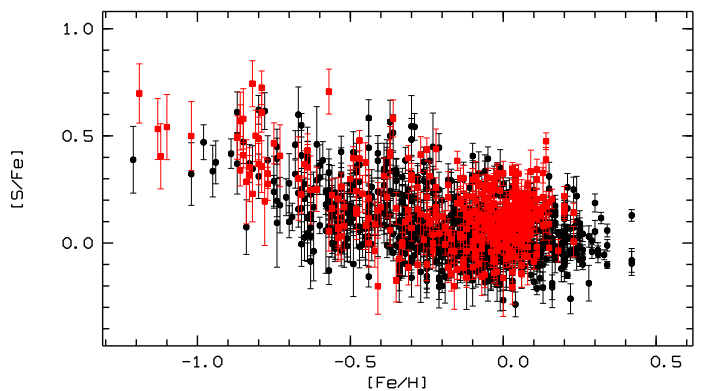}}
\end{center}
\caption[]{[S/Fe] vs. [Fe/H] for the sample stars including NLTE corrections.
Black and red symbols represent dwarf and giant stars, respectively,
and we see no difference in the two populations.
}
\label{plotsfe2}
\end{figure}

Figure\,\ref{plotsfe} depicts the 
 [Ca/Fe] versus [Fe/H] (provided by the iDR\,4 of the GES),
which is a typical and well studied $\alpha$-element, and of sulphur.
Figure\,\ref{plotsfe2} is the same as the top panel of Fig.~\ref{plotsfe} but includes the proper NLTE corrections
 we computed with the model atom described in \citet{korotin08,korotin09}, 
which are small and negative; the maximum absolute value is $-0.071$ and on average $\langle {\rm NLTE_{cor}}\rangle =-0.023 \pm 0.011$.
As expected, [S/Fe] is close to zero for solar-metallicity stars,
and it increases as the metallicity is reduced.

The absolute value of NLTE corrections increases as the temperature
increases achieving the maximum value at $\Teff= 6000-6500$\,K.
Corrections increase as well when decreasing gravity and can
exceed 0.1\,dex. Since our program giants have rather low
temperatures, the NLTE corrections are small in this case.

\citet{ecmult3} investigated granulation effects for the \ion{S}{i} 675.6\,nm triplet in few
solar-metallicity stars and found that the effects are small.

\begin{figure}
\begin{center}
\resizebox{\hsize}{!}{\includegraphics[draft = \draftflag, clip=true]
{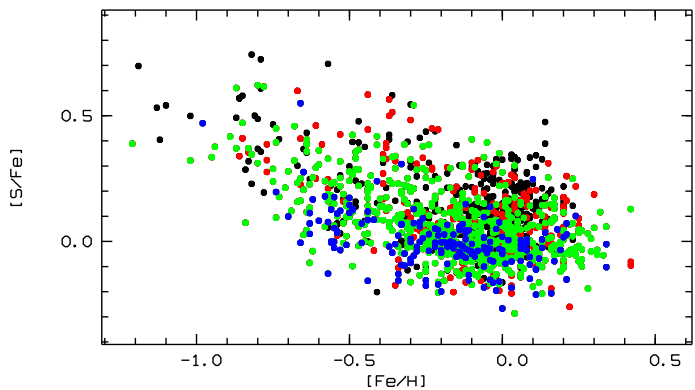}}
\end{center}
\caption[]{[S/Fe] vs. [Fe/H] for the complete sample of stars:
black, stars cooler than 5000\,K; red 
$5000\le$\teff $<5500$\,K; green 
$5500\le$\teff $<6000$\,K; blue hotter than 6000\,K.
}
\label{tcut}
\end{figure}

We have noticed that cool stars around solar metallicity tend to give a higher [S/Fe] than hotter ones.
This is summarised in Fig.\,\ref{tcut} where stars in different ranges in \teff\ are shown
with different colours. On average, the lower the temperature, the higher the [S/Fe].
We think this may be related to the CN lines that could be not properly taken into account in the wavelength
range of the \ion{S}{i} feature in the GES line-list.
This effect can have an impact on the analysis of the open clusters in which mainly cool {dwarfs and} giants 
have been observed.

At odds with the results presented by \citet[][]{nidever14,hayden15} from APOGEE data, 
our data show no indication of two different sequences in the [$\alpha$/Fe] vs. [Fe/H] planes 
for sulphur or calcium (see Fig.\,\ref{plotsfe}). The APOGEE samples and ours may cover overlapping regions 
in terms of galactocentric  distances and heights over the galactic plane (see Fig.\,\ref{fig_hg} and \ref{plot4panels}). 
A detailed comparison is, however, not possible. In fact, the trends detected in the APOGEE samples 
make use of a [$\alpha$/Fe] parameter obtained by combining measures of O, Mg, Si, Ca, Ti and S. 
Calcium and sulphur abundances presented by \citet[][see their Fig.~14]{holtzman15} show trends 
qualitatively similar to ours. Additionally, the APOGEE samples sizes are significantly larger than ours.

In the sample of stars we analysed there are some that have been 
observed as being members of globular clusters (54 stars)
and of open clusters (312 stars).
We verified the membership comparing radial velocities and metallicities. {For the young clusters we considered the members identified by \citet{spinainprep}, which were selected based on surface gravity and lithium line strength.}
We could derive the sulphur abundance in 21 clusters, 20 of which were analysed for the first time, as well as 47\,Tucanae, for which we already had
a sulphur determination \citep{sbordone09}. A more detailed discussion on this cluster is presented in Sect.\,\ref{47tuc}.
Table\,\ref{irpline2} provides S and Zn abundances for each cluster.

\begin{figure}
\begin{center}
\resizebox{\hsize}{!}{\includegraphics[draft = \draftflag, clip=true]
{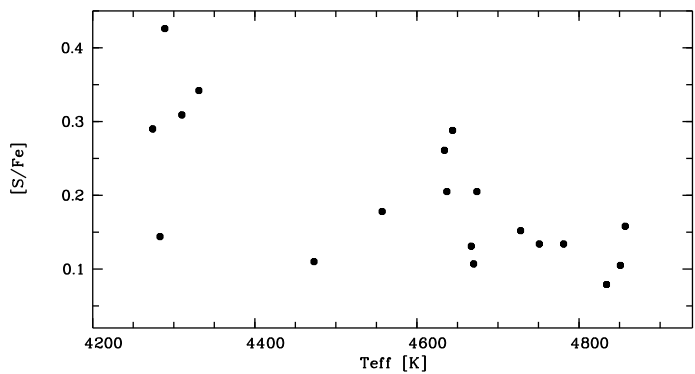}}
\end{center}
\caption[]{[S/Fe] versus effective temperature for the 19 stars in NGC\,6705.
}
\label{ngc6705trend}
\end{figure}

For some open clusters (e.g. Trumpler\,20, Trumpler\,23, NGC\,6705, Berkeley\,81) the high [S/Fe] could be explained
by the low \teff of the member stars.
This is in line with what was found in M\,67 for which, when selecting only the stars with \teff $>5000$\,K, 
we have a smaller star-to-star scatter, ${\rm [S/Fe]}=+0.0\pm 0.02$.
Also the 19 giant members of NGC\,6705 show a clear trend of increasing [S/Fe] by decreasing the stellar effective temperature.
The effect is evident in Fig.\,\ref{ngc6705trend}.
However, this cannot explain the high [S/Fe]=+0.21 of NGC\,2516 from two relatively warm stars (\teff $>5500$\,K).

\begin{figure}
\begin{center}
\resizebox{\hsize}{!}{\includegraphics[draft = \draftflag, clip=true]
{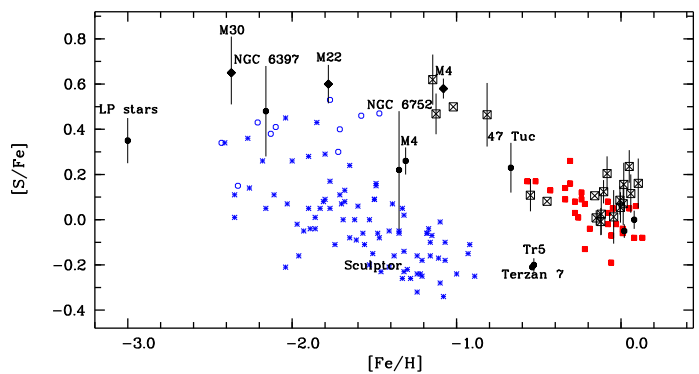}}
\end{center}
\caption[]{[S/Fe] vs. [Fe/H] for the stars members of open and globular clusters,
compared to analysis of clusters and field stars available in the literature. 
This is an update of figure\,4 of \citet{caffau14}, see references therein. 
For added samples: crossed black squares are from this work, blue stars are stars belonging to Sculptor dSph
from \citet{skuladottir15}, open blue circles are field stars from \citet{jonsson11},
and red filled squares are field stars from \citet{ecuvillon04}.
}
\label{plotcl2}
\end{figure}

In Fig.\,\ref{plotcl2} we compare the [S/Fe] versus [Fe/H] relation from the literature for open and 
globular clusters to our measurements after NLTE corrections are applied to our sulphur values. This figure is an update from our work in \citet{caffau14}. 

\begin{table*}
\centering
\scriptsize
\caption{\bf Distances, [Fe/H], [S/Fe] and [Zn/Fe] for stars belonging to clusters.}
\label{irpline2}
\begin{tabular}{lllllllllrrl}
\hline\hline
\noalign{\smallskip}
Cluster          & Type  & R$_{\sun}$ & R$_{\rm GC}$ & [Fe/H] & [S/Fe] & [S/Fe]$_{\rm NLTE}$ & [Zn/Fe] & [Zn/Fe]$_{\rm NLTE}$ & \multicolumn{2}{c}{Stars} & Comment\\
\cline{10-11}
 \noalign{\smallskip}
                 &   & kpc & kpc &    &    &        & & & N & Type                 & \\
\noalign{\smallskip}
\hline
\noalign{\smallskip}
NGC\,104/47\,Tuc & GC &  4.5   & 7.3  & $-0.81\pm 0.04$   & $+0.51\pm 0.14$ & $+0.46\pm 0.14$ & $+0.18\pm 0.11$ & $+0.17\pm 0.10$ & 19 & G   &  \\ %
NGC\,1851        & GC & 12.1   & 16.6 & $-1.13\pm 0.01$   & $+0.52\pm 0.09$ & $+0.47\pm 0.09$ & $+0.23\pm 0.01$ & $+0.20\pm 0.01$ & 2  & G   &  \\  %
NGC\,2808        & GC &  9.6   & 11.1 & $-1.15\pm 0.06$   & $+0.68\pm 0.11$ & $+0.62\pm 0.11$ & $+0.06\pm 0.01$ & $+0.05\pm 0.01$ & 2  & G   & high S \\  %
M\,15            & GC &  10.4  & 10.4 & $-2.65$           & $+1.75$         & $+1.68$         & $+0.16$         & $+0.17$         & 1  &     &  \\  %
NGC\,362         & GC & 8.7    & 9.40 & $-1.02$           & $+0.55$         & $+0.50$         & $-0.01$         & $-0.01$         & 1  & G   & high S \\  %
M\,67            & OC & 0.775  & 8.49 & $-0.12\pm 0.04$   & $+0.02\pm 0.06$ & $-0.01\pm 0.06$ & $-0.03\pm 0.08$ & $-0.06\pm 0.08$ & 18 & D,G & G with high S \\ %
NGC\,2243        & OC & 3.45   & 10.1 & $-0.55\pm 0.04$   & $+0.15\pm 0.07$ & $+0.11\pm 0.07$ & $+0.04\pm 0.05$ & $-0.00\pm 0.05$ & 9  & G   &  \\ %
Berkeley\,25     & OC & 9.1    & 15.6 & $-0.45$           & $+0.12$         & $+0.08$         & $+0.12$         & $+0.06$         & 1  & G   &  \\  %
NGC\,2451A       & OC & 0.2    & 8.0  & $-0.05\pm 0.06$   & $+0.03\pm 0.11$ & $+0.01\pm 0.12$ & $-0.02\pm 0.08$ & $-0.04\pm 0.08$ & 2  & D   &  \\ %
NGC\,2516        & OC & 0.35   & 7.93 & $-0.09\pm 0.04$   & $+0.21\pm 0.08$ & $+0.20\pm 0.08$ & $+0.48\pm 0.10$ & $+0.46\pm 0.10$ & 2  & D   & high S \\  %
NGC\,2547        & OC & 0.36   & 7.98 & $-0.12\pm 0.03$   & $+0.05\pm 0.09$ & $+0.02\pm 0.09$ & $+0.04\pm 0.04$ & $+0.02\pm 0.04$ & 2  & D   &  \\  %
IC\,2391         & OC & 0.15   & 7.94 & $-0.16$           & $+0.13$         & $+0.11$         & $+0.12$         & $+0.10$         & 1  & D   &  \\  %
Trumpler\,20     & OC & 3.0    & 6.9  & $+0.01\pm 0.05$   & $+0.10\pm 0.09$ & $+0.07\pm 0.09$ & $-0.27\pm 0.12$ & $-0.34\pm 0.12$ & 39 & G   &  \\ %
NGC\,4815        & OC & 2.5    & 6.9  & $-0.11\pm 0.02$   & $+0.16\pm 0.05$ & $+0.12\pm 0.05$ & $-0.23\pm 0.16$ & $-0.31\pm 0.16$ & 5  & G   & high S \\  %
Pismis\,18       & OC & 2.2    & 6.8  & $-0.01\pm 0.03$   & $+0.12\pm 0.06$ & $+0.09\pm 0.06$ & $-0.28\pm 0.11$ & $-0.35\pm 0.11$ & 6  & G   &  \\  %
NGC\,6005        & OC & 2.7    & 5.9  & $+0.06\pm 0.02$   & $+0.14\pm 0.09$ & $+0.12\pm 0.09$ & $-0.31\pm 0.09$ & $-0.37\pm 0.09$ & 12 & G   &  \\ %
Trumpler\,23     & OC & 2.0    & 6.3  & $+0.05\pm 0.04$   & $+0.27\pm 0.07$ & $+0.24\pm 0.07$ & $-0.21\pm 0.19$ & $-0.27\pm 0.18$ & 10 & G   & high S \\ %
NGC\,6633        & OC & 0.38   & 7.64 & $-0.15\pm 0.06$   & $+0.04\pm 0.04$ & $+0.01\pm 0.03$ & $-0.09\pm 0.05$ & $-0.12\pm 0.08$ & 8  & D,G &  \\  %
NGC\,6705        & OC & 1.9    & 6.3  & $+0.02\pm 0.06$   & $+0.20\pm 0.10$ & $+0.16\pm 0.10$ & $-0.28\pm 0.18$ & $-0.37\pm 0.17$ & 19 & G   & trend S with \teff \\ %
Berkeley\,81     & OC & 3.0    & 5.7  & $+0.10\pm 0.06$   & $+0.19\pm 0.11$ & $+0.16\pm 0.11$ & $-0.36\pm 0.16$ & $-0.42\pm 0.15$ & 6  & G   & trend S with \teff \\  %
NGC\,6802        & OC & 1.8    & 7.1  & $-0.00\pm 0.02$   & $+0.09\pm 0.07$ & $+0.05\pm 0.07$ & $-0.15\pm 0.13$ & $-0.22\pm 0.13$ & 9  & G   &  \\  %
\noalign{\smallskip}
\hline
\end{tabular}
\smallskip
\tablefoot{GCs distances from \citet{harris96} (2010 edition). For OCs: distance for NGC\,2243, \citet{bragaglia06}; 
Be\,25, \citet{carraro07}; M\,67, \citet{montgomery93}; all the others from \citet[][and reference therein]{spinainprep}.
 Galactocentric radii have been computed assuming a distance of the Sun to the Galactic center of 7.94 kpc \citep{eisenhauer}.
{[Fe/H] is derived averaging among the selected stars in each cluster.} The solar abundances here applied are those of the third column in Table\,\ref{solar_table}.
{Average [Fe/H] values for the open clusters are by about 0.1 dex below those reported by \citet{jacobson16} and \citet{spinainprep}, due to the use of different solar reference values, and to the inclusion only of the subset of stars with measured sulphur.}
}
\end{table*}

\subsection{A possible S - Na correlation in 47\,Tucanae}\label{47tuc}
Sulphur abundances in \object{NGC 104} (\object{47 Tuc}) were investigated by \citet{sbordone09}.
They determined sulphur abundances in 4 turn-off and 5 subgiant stars, using VLT-UVES spectra and
measuring lines of \ion{S}{i} Mult.\,1 around 922 nm. They claimed a statistically significant
{\em positive correlation} of [S/Fe] with [Na/Fe], which, if confirmed, would be of high
interest in the context of the investigation of multiple populations in globular clusters, especially
because there is no obvious mechanism of sulphur production as part of any currently considered
globular cluster self-enrichment mechanisms \citep{sbordone09}.

\begin{figure}
\begin{center}
\resizebox{\hsize}{!}{\includegraphics[draft = \draftflag, clip=true]
{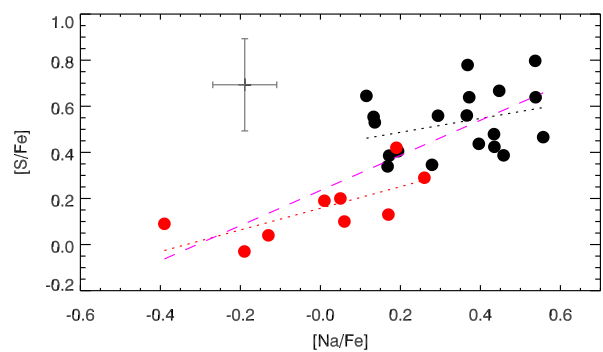}}
\end{center}
\caption[]{[S/Fe] vs. [Na/Fe] for stars in \object{NGC 104}. Black points are GES measurements in giants, 
red points are \citet{sbordone09} TO and SGB stars,
shifted to our adopted solar abundance scale. Linear fits are included, the magenta line indicating the fit 
to the whole sample. A conservative error estimate is also shown.
}
\label{S_Na_47tuc}
\end{figure}

In Fig. \ref{S_Na_47tuc} we plot [S/Fe] vs. [Na/Fe] in \object{NGC 104} for the \citet{sbordone09} sample
together with the current one. The \citet{sbordone09} abundance ratios have been brought into the same scale
used in GES by accounting for the slightly different assumed solar abundances. 

When looking at the \citet{sbordone09} and GES sample separately, neither produces a significant slope 
when fitted linearly (GES, $0.30\pm0.34$, \citealt{sbordone09} $0.47\pm0.37$). 
The \citet{sbordone09} sample showed a very high likelihood of correlation between [S/Fe] and [Na/Fe] 
through a Kendall $\tau$ test. However, when the two samples are {\em taken together} the slope of the linear 
fit is highly significant ($0.76\pm0.18$). Also, it is remarkable how the two samples cover different ranges in [Na/Fe]. 
The GES sample appears to cover a range consistent with recent [Na/Fe] measurements in \object{NGC 104} giants 
(\citealt{cordero14}, \citealt{thygesen14}, \citealt{johnson15}), while analyses of TO and SGB stars find a distribution 
significantly more extended towards low Na abundances (\citealt{dobrovolskas14}, \citealt{marino16}). 
It is particularly intriguing that when stars belonging to the bright and faint SGB of 47 Tuc (bSGB, fSGB) 
are discriminated in \citet{marino16}, the bSGB stars appear Na-poor, and match the range in Na observed 
by \citet{dobrovolskas14} and \citet{sbordone09}. 
The fSGB stars, on the other hand, match the Na abundances covered in the GES sample and other RGB-based studies. 
On the one hand, the bSGB is more populated and brighter so it is reasonable that the \citet{sbordone09} 
and \citet{dobrovolskas14} samples, designed to study the weak \ion{Li}{i} 670.8nm doublet, were drawn from this population. 
On the other hand, it is unclear why RGB samples appear to lack the Na-poor tail detected in the prominent bSGB. 
Investigating the [Na/Fe] distributions of different \object{NGC 104} populations is outside the scope of the present paper, 
but the similarity between the [Na/Fe] distribution in the two \object{NGC 104} SGBs, 
and the two samples here investigated for sulphur and zinc, suggest the possibility that the \citet{sbordone09} 
and the GES samples might be drawn from different subpopulations of the cluster, possibly each without internal [S/Fe] spread, 
but rather characterised by different [S/Fe] values. In this case, once the \citet{sbordone09} sample is brought to the 
solar abundance scale employed in our analysis, they would correspond to [S/Fe]=$0.16\pm0.14$ 
and [S/Fe]=$0.53\pm0.13$ (\citealt{sbordone09}, GES).

Caution in the comparison is in order since different multiplets are used, as well as stars of different 
atmospheric parameters, so systematic differences between the two samples might be induced if line formation 
systematics (3D, NLTE...) are not correctly accounted for. 
However, the currently available data indicate that \object{NGC 104} displays either a spread in [S/Fe], 
strongly correlating with [Na/Fe], or two subpopulations characterised by significantly different values of [S/Fe].

\subsection{Zinc over iron}

\begin{figure}
\begin{center}
\resizebox{\hsize}{!}{\includegraphics[draft = \draftflag, clip=true]
{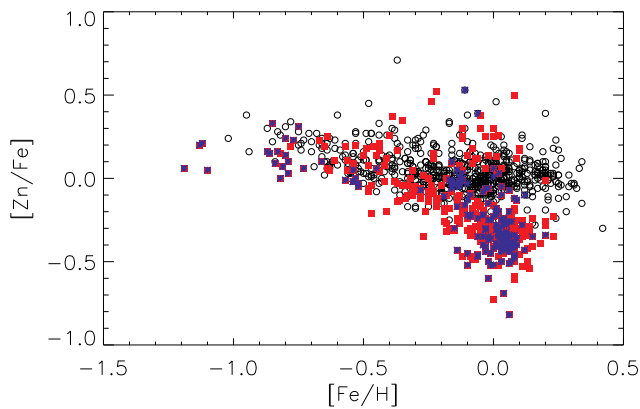}}
\end{center}
\caption[]{[Zn/Fe] vs. [Fe/H] for the sample stars analysed for sulphur. 
Dwarf stars are represented as black open circles, giant stars as red 
squares, and stars in clusters as blue crosses.
}
\label{plotznfe}
\end{figure}

In Fig.\,\ref{plotznfe} we show the [Zn/Fe] versus [Fe/H] for the sample of 
stars analysed for sulphur. A large scatter in [Zn/Fe] is evident overall 
around solar metallicity.
When we divide the sample into dwarf stars ($\log{\rm g}>3.45$)
and giant stars we realise that the large spread is mainly due to giants. The 897 
stars classified as dwarfs give $\langle{\rm [Zn/Fe]_d}\rangle = { 0.07\pm 0.11} $
while the 404 giants give $\langle{\rm [Zn/Fe]_g}\rangle = -0.12\pm 0.22$. 
Hence, the scatter of [Zn/Fe] for giant stars is larger than both the observational 
error, which is on average $\approx 0.12$ (Fig.\,\ref{plotznfe}), and the dispersion observed 
within dwarf stars. On average, giants show lower [Zn/Fe] than dwarfs. 
Such differences are even more evident when we select the 525 stars around solar 
metallicity, $-0.1<{\rm [Fe/H]}<+0.1$, for which we derive $\langle{\rm [Zn/Fe]}
\rangle = -0.06\pm 0.19$. 
In this case the 295 dwarfs contribute with $\langle{\rm [Zn/Fe]_d}\rangle = 0.04
\pm 0.10$ and the 230 giants with $\langle{\rm [Zn/Fe]_g}\rangle =-0.20\pm 0.20$. 
The same calculations for the 162 stars (mainly giants) that are members of 
clusters provide $\langle{\rm [Zn/Fe]_c}\rangle = -0.19\pm 0.22$ for the overall 
sample, and $\langle{\rm [Zn/Fe]}\rangle = -0.20\pm 0.20$ for stars around solar 
metallicity. Stars in clusters contribute to lower the average [Zn/Fe] of the 
giant sample at [Fe/H]$\approx 0.0$ but apparently they are not the main drivers 
of the large scatter, as can be seen in Fig.\,\ref{plotznfe}.

To investigate if the differences between dwarfs and giants are real or if 
the observed trend is driven by analysis and/or observational biases we made 
several tests. 

\begin{figure}
\begin{center}
\resizebox{\hsize}{!}{\includegraphics[draft = \draftflag, clip=true]
{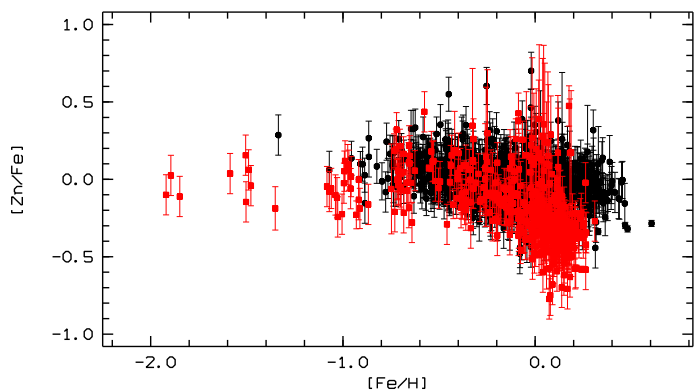}}
\end{center}
\caption[]{[Zn/Fe] vs. [Fe/H] for the complete GES sample stars with detection of zinc.
Black filled circles represent dwarf stars, red filled squares represent giant stars.
}
\label{plotznfeall}
\end{figure}

First, we checked if this behaviour is a consequence of restricting the GES 
sample to stars with detected sulphur. This is shown in Fig.\,\ref{plotznfeall}, 
where we plot [Zn/Fe] vs [Fe/H] for the complete sample of 1724 stars. In the 
solar-metallicity regime we can see that both the low [Zn/Fe] values and the large 
scatter are clearly reproduced in the case of giant stars.

Second, we tested possible biases due to NLTE effects. Using the computations 
of \citet{takeda05} we could derive for our sample a NLTE correction of 
$-0.06\pm 0.02$ for zinc, which is well within the observational uncertainty.
Furthermore the correction goes in the opposite direction, i.e. it is negative,
meaning that it further decreases the [Zn/Fe] value. For iron, we provide an 
estimate of the NLTE effects using the results of \citet[][see also \citealt{lind12}]{mashonkina08}.
According to their calculations the NLTE correction,
mainly affects the \ion{Fe}{i} lines, is smaller than $0.1$\,dex for both giants and 
dwarfs at solar metallicity, and it is always positive. Thus, also in this 
case the NLTE effect is expected to be negligible and to depress the [Zn/Fe] 
values further on. 

Third, to investigate the effects that granulations can have on the abundances
we computed zinc abundances for 22 hydrodynamical models and their reference ${\rm 1D}_{\rm LHD}$
models from the CIFIST grid \citep{cifistgrid}, for two metallicities (0.0 and $-1.0$)
and we computed the 3D correction as in \citet{zolfito}.
To study the effects in dwarf stars, we selected the solar model and three
effective temperatures (5500, 5900 and 6250\,K) for two gravities (4.0 and 4.5).
In all cases but one (for the 481\,nm line and the hottest model at \logg =4.0 
and [Fe/H]=0.0), the 3D corrections are positive, on average $0.08$ and $0.06$
for the 481\,nm and 636\,nm line respectively.
For the giant stars we investigated three models at 5000\,K (gravity of 2.5, 3.0 and 3.5)
and two at 4500\,K (gravity 2.0 and 2.5).
The 3D corrections are slightly smaller for giants, $0.04$ and $0.02$
for the 481\,nm and 636\,nm line respectively.
Hence, also the granulation effects are also comparable to the uncertainties.

\begin{figure}
\begin{center}
\resizebox{\hsize}{!}{\includegraphics[draft = \draftflag, clip=true]
{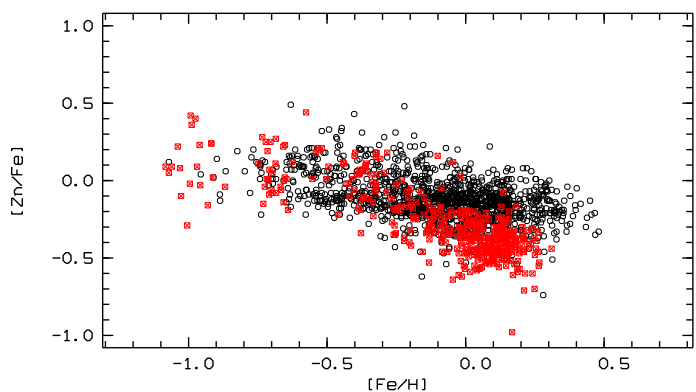}}
\end{center}
\caption[]{[Zn/Fe] vs. [Fe/H] in the case of zinc derived with \mygi\
only for the 481.0\,nm line.
Symbols are as in Fig.\,\ref{plotznfe}.
}
\label{znmygi}
\end{figure}

Fourth, we considered the zinc abundances derived by \mygi\ when fitting 
the strongest of the two \ion{Zn}{i} lines at 481.0\,nm. 
In principle, if there was a problem in the abundances derived from the 636.2\,nm line,
either due to the blending CN lines or to the \ion{Ca}{i} 636.1\,nm auto-ionisation
line, we may expect a systematic difference.
The result is shown in 
Fig.\,\ref{znmygi}. We notice that, although the star-to-star scatter is smaller 
than the homogenised A(Zn) values (by about 0.08\,dex at metal-rich regime), the different behaviour 
of dwarf and giants is still there, and in particular around solar-metallicity,
giant stars show much lower [Zn/Fe] values than dwarfs. Furthermore, opposite 
to the case of sulphur (Fig.\,\ref{ngc6705trend}), a clear trend of increasing/decreasing zinc 
abundance as a function of the effective temperature is not evident. 
We also selected giant stars, similar in stellar parameters but with a difference of
at least 0.4\,dex in [Zn/Fe], we compared the observed spectra and checked the
results we obtained from \mygi. The spectra of low- and high-Zn abundance
stars appear similar and \mygi\ provides very similar values, close to the 
low measurement. This test leads us to have confidence in the presence of a low-Zn abundance
population of giant stars, while casting doubt on the high values provided in iDR4.
This conclusion is reinforced by the absence of high-Zn abundance giants in Fig.\,\ref{znmygi} 
when compared to Fig.\,\ref{plotznfeall}. 
We suspect that the high [Zn/Fe] values are the result of an incorrect synthesis
of the region around the \ion{Zn}{i}  636.2\,nm line in iDR4.

{Finally, to investigate the impact of the uncertainties in the stellar
parameters on [Zn/Fe], we took into consideration a giant star with a low-Zn abundance.
With \mygi\ we derived the Fe and Zn abundances by changing effective temperature,
gravity and micro-turbulence according to their uncertainties.
A change in $\pm 110$\,K in \teff\ implies a change in [Zn/Fe] by $^{-0.07}_{+0.08}$;
by changing \logg\ by $\pm 0.22$ the change in [Zn/Fe] is of $\pm{0.03}$;
a change in the micro-turbulence of $\pm 0.10$\kms\ implies a change in [Zn/Fe] of about $\mp 0.02$.
All the changes in [Zn/Fe] are too small to alter the above-described picture.
}

\begin{figure}
\begin{center}
\resizebox{\hsize}{!}{\includegraphics[draft = \draftflag, clip=true]
{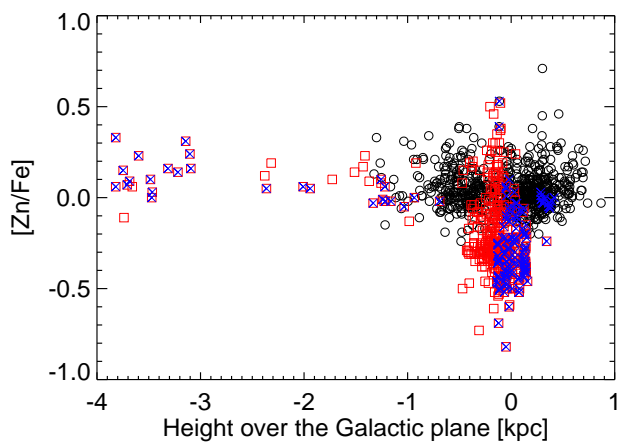}}
\end{center}
\caption[]{[Zn/Fe] as a function of the height above (or below) the Galactic plane.
Symbols as in Fig.\,\ref{plotznfe}.
}
\label{fig_hg}
\end{figure}

In conclusion, after all these tests we believe it is unlikely that the difference 
in [Zn/Fe] between dwarfs and giants is due to some systematic error in the analysis.
We therefore address the question if the two samples come from the same parent population. 

\subsection{Different populations, and radial gradient in [Zn/Fe]?}

The selection function specific to the GES UVES targets \citep{smiljanic14} 
results in a magnitude-limited sample primarily aimed at local FGK dwarfs, while 
the (less numerous) giants generally reside at much larger distances from the Sun. 
The giant stars are also targeted on purpose with UVES in the bulge fields and 
in the CoRoT fields. In fact, in the giant sample 20\% are in the bulge direction 
(actually inner disk giants) and 10 \% in the CoRoT fields. 
This combined with the open clusters fully explain why the sample is strongly concentrated on the galactic plane.
For the clusters we took the distances from the literature (see Table\,\ref{irpline2}).
For the field stars, distance moduli are computed using a Bayesian method on the Padova isochrones 
\citep[CMD 2.7]{bressan12} and using the magnitude independent of extinction 
$K_{\rm J-K} = K - \frac{A_K}{A_J - A_K} (J-K)$ with extinction coefficients computed applying 
the \cite{FitzpatrickMassa07} extinction curve on the \citet{CastelliKurucz03} SEDs. 
The prior on the mass distribution used the IMF of \cite{chabrier01} while the prior on age was chosen flat. 
Stars too far from the isochrones were rejected using the $\chi^2_{0.99}$ criterion.
Moreover, giants were targeted predominantly in open clusters, or in globular clusters 
that were observed as calibrators \citep{pancino16}. The dwarf sample happens to be 
entirely located within $\leq 1.5$\,kpc from the Sun, with a peak of the distribution 
at $D\approx 0.5$\,kpc.
Giant stars cover a 
much larger range in distance, $0.0 < D/{\rm kpc} < 16$, with all stars, but one  
further than 6\,kpc from the Sun residing in clusters. The distant giants sample is 
all at low metallicity ([Fe/H]$\le -0.5$), and almost entirely hosted in globular clusters. 
Local giants are predominantly metal-rich, and largely hosted in open clusters. 

In Fig.\,\ref{fig_hg} we show the [Zn/Fe] ratio as a function
of the height of the stars from the Galactic plane.
It is then obvious that the giant sample is more strongly concentrated
on the Galactic plane, while dwarfs are mostly observed at heights
larger than 0.5\,kpc from the plane.
If we select all dwarfs and giants with heights $\le 1$\,kpc
from the plane then the distribution of [Fe/H] of the two
samples is distinctly different. Both dwarfs and giants have a peak
at solar metallicity, yet the dwarfs have a large excess of lower
metallicity stars down to [Fe/H]=--1.0.
This suggests that our dwarf stars sample is largely dominated
by the thick disc objects, while the field giants and the open clusters are
an almost pure thin disc population.

On average, giants have been observed with UVES in the Gaia-ESO Survey 
at Galactic latitudes lower than dwarfs: $\left<|b|\right>_G \approx 9^\circ$, 
while dwarfs $\left<|b|\right>_D \approx 30^\circ$. This bias is not compensated 
by the geometrical bias: for a given apparent magnitude, giants are more 
distant, so that they are observed at larger heights from the Galactic 
plane than dwarfs of the same apparent magnitude and Galactic latitude. 
Of course the distinction between thin and thick disc based only on the 
height from the Galactic plane is very crude. The second Gaia data release 
will provide parallaxes and proper motions for all these stars. 
When coupled with our radial velocities we shall be able to compute Galactic 
orbits for all these stars and classify them as belonging to the thin or the 
thick disc.

\begin{figure*}
\begin{center}
\resizebox{\hsize}{!}{\includegraphics[draft = \draftflag, clip=true]
{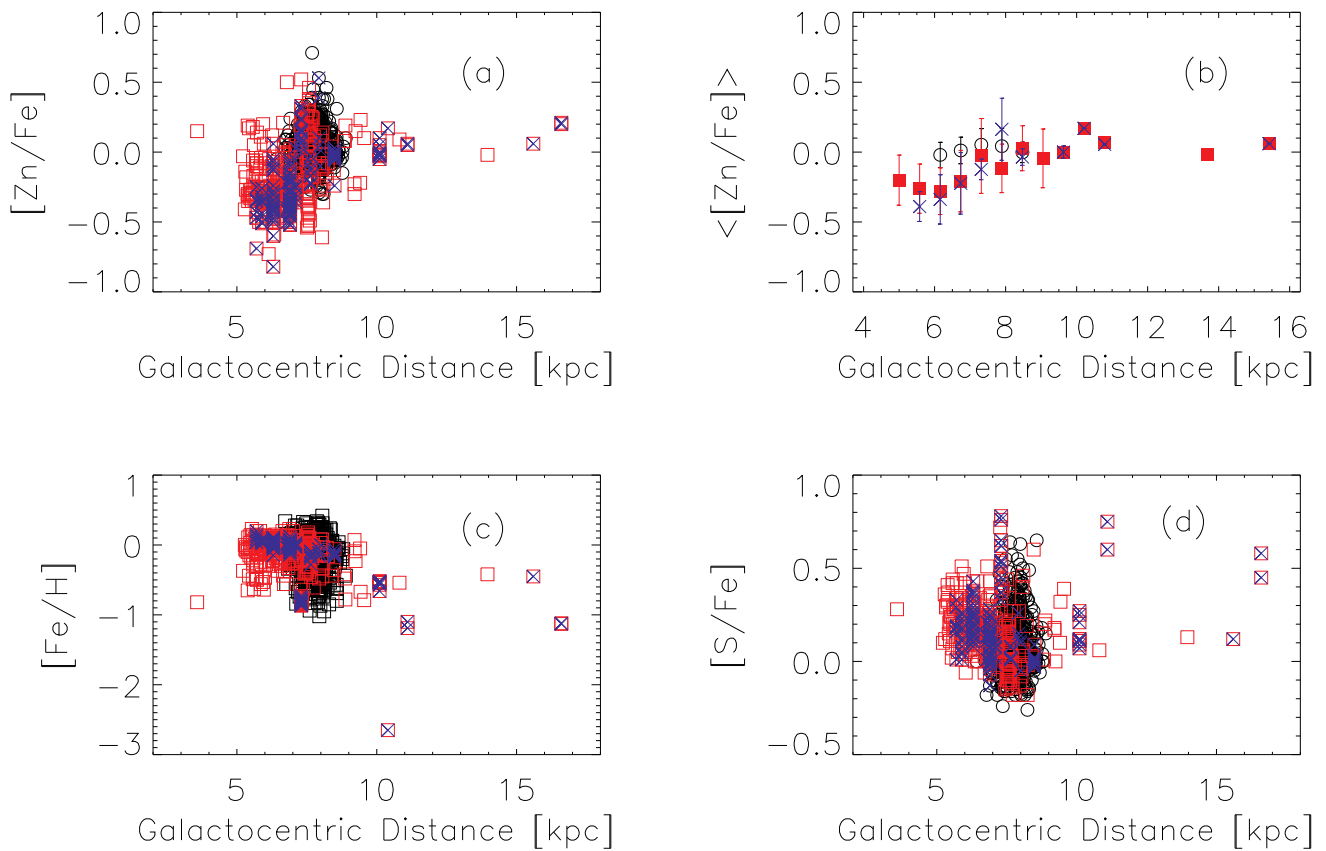}}
\end{center}
\caption[]{{\it Panels a, c, d:} The [Zn/Fe] (a), the [Fe/H] (c), and the 
[S/Fe] abundances (d) for the sample of stars analysed for sulphur as a 
function of their Galactocentric distance. Symbols are the same of 
Fig.\,\ref{plotznfe}. {\it Panel b:} The average [Zn/Fe] in different 
distance bins for dwarfs (black circles), giants (red squares) and for 
stars in clusters (blue crosses).}
\label{plot4panels}
\end{figure*}

Following the findings of \cite{fuhrmann98,fuhrmann99,fuhrmann04}, that 
have been verified by subsequent investigations \citep[see, e.g.][and 
references therein]{wojno,haywood,mikolaitis,GES14}, one expects (kinematically selected)
thin disc stars to have lower $\alpha$-to-iron ratios than thick disc 
stars of the same metallicities. If our dwarf stars sample is dominated 
by the thick disc stars, as suggested by Fig.\,\ref{fig_hg}, we would expect 
higher $\alpha$-to-iron ratios than in the giant star sample. However, 
as can be appreciated in Fig.\,\ref{plotsfe}, there is no clear distinction 
between dwarfs and giants in the $\alpha$-to-iron ratios. 

\begin{figure}
\begin{center}
\resizebox{\hsize}{!}{\includegraphics[draft = \draftflag, clip=true]{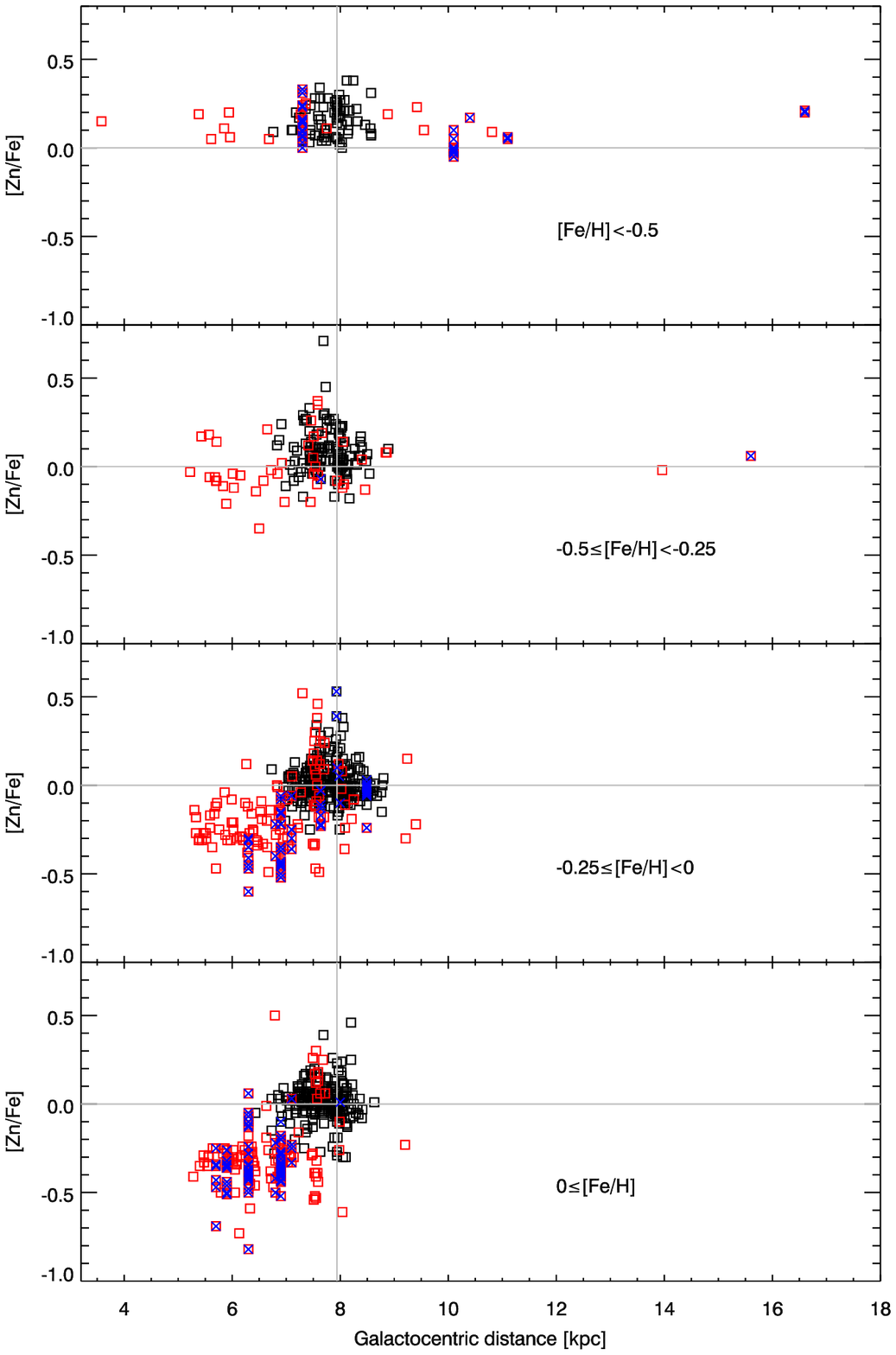}}
\end{center}
\caption[]{[Zn/Fe] plotted against R$_{\rm GC}$ for dwarfs (black symbols) and giants (red symbols). Blue crosses indicate stars in clusters. The plot is split to contain only stars in the indicated metallicity range.}
\label{ZnRgcBin}
\end{figure}

In Fig.\,\ref{plot4panels} we plot chemical abundances of giants, dwarfs, 
and stars in clusters as a function of their distance (from Table\,\ref{irpline2}) from the Galactic centre.
In panels (a), (c) and (d) of Fig.\,\ref{plot4panels}, we can see the dwarfs' sample (black open circles) located at around 8\,kpc 
from the Galactic centre (where the Sun is situated), consistent in [Fe/H] and [Zn/H] with disc stars.
Giant stars belonging to clusters (globular or open) 
are depicted in panels (a), (c) and (d) of Fig.\,\ref{plot4panels} as red squares with blue crosses.
In panel (b) we binned the abundances in various distance bins. 
For Galactocentric distances larger than  7.5\,kpc, on average 
$\langle$[Zn/Fe]$\rangle\approx 0.0$, with a good agreement between field giants, giant stars in clusters
and dwarf stars. As we move to smaller distances from the
Galactic centre the giants and the cluster stars display a smaller
$\langle$[Zn/Fe]$\rangle\approx -0.2$, while the dwarf stars
remain at $\langle$[Zn/Fe]$\rangle\approx 0.0$.

However, panels (a) and (b) in Fig.\,\ref{plot4panels} do not allow to disentangle the superimposed effects of metallicity and galactocentric
distance. For this purpose, in Fig. \ref{ZnRgcBin} we plot [Zn/Fe] versus the galactocentric radius, but splitting the whole sample in four metallicity
ranges. Once this is done, a few general behaviors of the sample appear with more clarity:
\begin{itemize}
\item At low metallicity ([Fe/H]$< -0.5$) both dwarfs and giants show slightly super-solar [Zn/Fe], constant at all galactocentric radii.
\item As metallicity increases ($-0.5\le$[Fe/H]$<-0.25$, and $-0.25\le$[Fe/H]$<0$), sub-solar [Zn/Fe] giants appear {\em in the inner disk}. The average value for dwarfs decreases to [Zn/Fe]$\approx 0$. At the same time, {\em at the solar radius} there is little evidence of a discrepancy between dwarfs and giants. 
\item The most Zn-poor giants all appear at small galactocentric radii (R$_{\rm GC}<7$kpc and for high metallicities ([Fe/H]$>0$).
\end{itemize}
It thus appears that the dwarf-giants discrepancy in Fig. \ref{plotznfe} to \ref{plot4panels} is driven by the superposition different selection effects. 
On the one hand, giant stars appear to be much more concentrated on the plane, thus likely belonging preferentially to a younger population than the dwarfs. On the other hand, Zn-poor giants appear to prevalently belong to the inner disk, and all display solar, or supersolar metallicities. Although it cannot be excluded that a systematic difference in the analysis exists between dwarfs and giants above [Fe/H]$\approx 0$, due to the limited overlap in R$_{\rm GC}$ at high metallicity, the trend appears already quite evident in the $-0.25\le$[Fe/H]$<0$ bin, where there is a healthy sample of giants at higher galactocentric radii whose [Zn/Fe] is in agreement with the one of the dwarfs. {It is also worth noticing that there is a small subpopulation of dwarfs with low [Zn/Fe] but their small number (seven dwarfs with [Zn/Fe]$< -0.3$) prevents us from drawing any strong conclusion. These are prevalently cool dwarfs, hence faint ones, and closer to the galactic plane that the bulk of the dwarfs in our sample.} We are thus inclined to consider the low [Zn/Fe] ratios we observe as a real signature of the chemical enrichment of the inner MW disk.

{The complex behavior of [Zn/Fe] with R$_{\rm GC}$, metallicity, and age is shown in two recent investigations of high precision abundances in nearby solar twins. \citet{nissen15} analyzes  a sample of 21 solar twins 
and finds a clear trend in [Zn/Fe], which increases with increasing stellar age (by about 0.1\,dex over about 8\,Gyr).
A similar trend is found also for [Mg/Fe] and [Al/Fe]. Similar results are found by \citet{spina16} in a study of 9 objects. The much more heterogeneous, and lower quality GES sample cannot detect such subtle variations, and lacks precise age estimates: in Fig. \ref{ZnRgcBin}, the distribution of solar-metallicity, solar-galactocentric-radius dwarfs disperse in [Zn/Fe] by a value comparable to the extent of the correlation found by \citet{nissen15} (and Mg and Al have comparable dispersion). These works, however, support the finding of a dependency of [Zn/Fe] from the star formation epoch and environment that is in such stark display in our inner-disk giants.}

\subsection{A possible effect of SN Ia dilution?}
\label{dilution}

The concentration of low [Zn/Fe] stars on the galactic plane and at small galactocentric radii suggests they might represent a younger population than the dwarf sample, whose Zn would then be more diluted by SN Ia ejecta (which are believed to be almost Zn-free). 
However, said dilution would affect $\alpha$ elements as well to some extent. We thus proceeded to test  this scenario through a simple calculation. 
Inspecting Fig. \ref{plotsfe} and \ref{plotznfe}, one notices that at  A(Fe) = 7.22 (i.e. [Fe/H]$\approx-0.3$) 
giant and dwarf stars show on average the same ratios of alpha elements, in particular A(Ca)-A(Fe) and 
A(S)-A(Fe), along with (at or outside the solar circle) the same A(Zn)-A(Fe).
{ By remembering that:
\begin{multline*}
A({\rm Fe}) = \log \frac{N_{\rm Fe}}{N_{\rm H}} + 12 = \log \frac{N_{\rm Fe}\times m_{\rm Fe}}{N_{\rm H}\times m_{\rm H}} + 12 - \log \frac{m_{\rm Fe}}{m_{\rm H}}\\ \approx  \log \frac{N_{\rm Fe}\times m_{\rm Fe}}{M_{g}} + 12 - \log \frac{m_{\rm Fe}}{m_{\rm H}} \\
= \log \frac{M_{\rm Fe}}{M_{g}}+ 12 - \log \frac{m_{\rm Fe}}{m_{\rm H}}
\end{multline*}
}
where $m_{\rm Fe}$ ($m_{\rm H}$) is the atomic mass of iron (hydrogen), and that
\begin{equation*}
A({\rm X})-A({\rm Fe}) = \log \frac{M_{\rm X}}{M_{\rm Fe}} - \log \frac{m_{\rm X}}{m_{\rm Fe}}, 
\end{equation*}
we can compute the mass of each chemical element, $M_{\rm X}$, as a function of the mass 
of gas, $M_{g}$, out of which these stars have formed. For example, for $M_{g} \approx 10^{10} M_{\odot}$, 
we get $M_{\rm Fe} = 9.2 \times 10^{6} M_{\odot}$.

By assuming that the following chemical enrichment of the gas, which leads to $A({\rm Fe})^{\rm obs} = 7.52$
(i.e. [Fe/H]$^{\rm obs} \approx 0.0$) and $A({\rm Zn})-A({\rm Fe})]^{\rm obs} \approx -3.1$, is {\em only} driven by SN Ia, 
we can compute the required mass of SN Ia, $M_{\rm SN Ia}$, along with the final mass of gas, $M_{g}^{out}$, 
by using the following equations:
\begin{multline*}
A({\rm Fe})^{obs} = \log \frac{M_{\rm Fe} + Y^{\rm Fe_{SN Ia}}\times M_{\rm SN Ia}}{M_{g}^{out}} - \log \frac{m_{\rm Fe}}{m_{\rm H}} +12\\
\left(A({\rm Zn})-A({\rm Fe})\right)^{obs} = \log \frac{M_{\rm Zn} + Y^{\rm Zn_{\rm SN Ia}}\times M_{\rm SN Ia}}{M_{\rm Fe} + Y^{\rm Fe_{\rm SN Ia}}\times M_{\rm SN Ia} }- \log \frac{m_{\rm Zn}}{m_{\rm Fe}}
\end{multline*}
We can then exploit the derived $M_{\rm SN Ia}$ and $M_{g}^{out}$ values to get the expected (``out'') alpha-to-iron 
ratios of several chemical elements. By using the different SN Ia yield models from \citet{iwamoto99}, and averaging 
among the corresponding results, we get $\left(A({\rm Mg})-A({\rm Fe})\right)^{out} = 0.12 \pm 0.002$, 
$\left(A({\rm Ca})-A({\rm Fe})\right)^{out} = -1.14 \pm 0.04$ and $\left(A({\rm S})-A({\rm Fe})\right)^{out} 
= -0.28 \pm 0.03$\footnote{These findings are independent on the assumed mass of gas, $M_{g}$.}, 
which are in good agreement with the observed values at $A({\rm Fe})$ = 7.52, i.e. $\left(A({\rm Ca})-A({\rm Fe})\right)^{obs} = -1.16$, 
$\left(A({\rm S})-A({\rm Fe})\right)^{obs} = -0.28$, with the exception of Mg, for which $\left(A({\rm Mg})-A({\rm Fe})\right)^{obs} = 0.24$ 
is 0.12 dex higher than the theoretical expected value.
Thus, although this picture needs to be carefully tested against other chemical elements and by exploiting 
detailed cosmological chemical evolution models, we conclude that it is plausible. Clearly, if the enrichment 
of SN Ia is really at the origin of the low zinc to iron ratio observed in giant stars located in the inner thin disk,
the same trend should be observed in dwarf stars, once observed in the same region. Thus, it can in principle be
tested observationally, although 30m-class telescopes will likely be needed.

\subsection{Sulphur over zinc}

\begin{figure}
\begin{center}
\resizebox{\hsize}{!}{\includegraphics[draft = \draftflag, clip=true]
{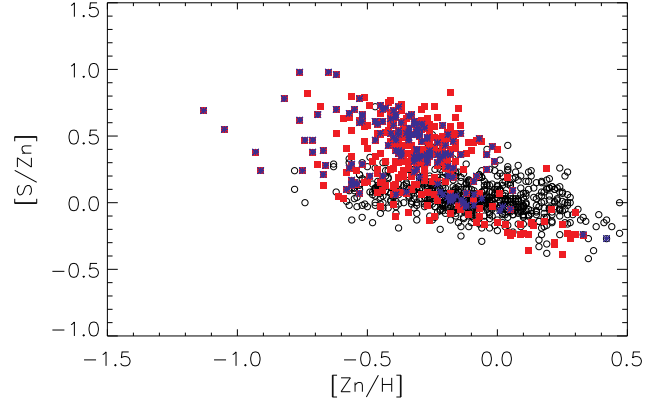}}
\end{center}
\caption[]{[S/Zn] vs. [Zn/H] for the sample of GES stars we analysed for sulphur.
Symbols are as in Fig.\,\ref{plotznfe}.
}
\label{plotszn}
\end{figure}

In Fig.\,\ref{plotszn} the [S/Zn] versus [Zn/H] abundances for the 
sample of stars analysed for sulphur are shown, by distinguishing
among dwarfs, giants and stars in clusters. 
As already noticed for the [Zn/Fe] vs [Fe/H] trend, 
we clearly see that dwarfs and giants behave very differently.
Dwarf stars show a constant [S/Zn] value (or perhaps a slight slope)
around [S/Zn]$\approx 0.0$ within [Zn/H]$\approx (-0.1,0.5)$. The slight slope in the
dwarf sample might simply be the effect of [S/Fe] increasing more rapidly than [Zn/Fe] as metallicity decreases.
On the other hand, giant stars show a declining
[S/Zn] trend with [Zn/H], with super-solar [S/Zn] values at [Zn/H]$<0.0$
and large scatter. This is due to the giant sample reaching deeper in the inner disk, 
where the metal rich population shows solar [S/Fe] but subsolar [Zn/Fe].

\section{Conclusions}

We  analysed the sulphur and zinc abundances in a large sample of Galactic stars. 
We here below summarise what we found on the analysed sample.

On sulphur:
\begin{itemize}
\item
Sulphur behaves as an $\alpha$-element, with a typical behaviour
of [S/Fe] compatible with 0.0 within uncertainties for stars around solar metallicities.
The values of [S/Fe] increase for decreasing metallicity with, at around
[Fe/H] of --1.0, a constant value of [S/Fe].
Unfortunately, due to the weak sulphur feature, the stars in our sample with 
${\rm [Fe/H]}\le -1.0$ are only seven, so that we cannot make conclusions on the behaviour
of sulphur in the metal-poor regime.
\item
With the line-list we used, we detect a clear trend of [S/Fe] as a function of $T\rm_{eff}$.
Further investigations on the contribution of CN molecules in the wavelength range will follow.
\item
We could not find a cluster with ``low'' [S/Fe], like Trumpler\,5 that, according to \citet{caffau14},
has a [S/Fe] compatible with Local Group galaxies \citep{caffau05b}.
All the clusters with ${\rm [Fe/H]}\le -0.4$ are enhanced in S, as are field stars.
\item
The open clusters around solar metallicity on average show high [S/Fe] values,
but we attribute it to the presence of cool stars whose sulphur abundances are systematically ``high''.
\item
We confirm and strengthen the detection \citep{sbordone09} of a significant [S/Fe] spread in \object{NGC 104}, 
which appears to correlate to a high degree of significance with [Na/Fe]. 
While at face value the data appear to show an actual trend of [S/Fe] with [Fe/H], we cannot rule out that 
we may actually be sampling two different \object{NGC 104} populations, one S-rich and one S-poor, but each without internal sulphur spread.
\end{itemize}

On zinc:

\begin{itemize}
\item In the GES sample, 
there is a sizeable scatter in the [Zn/Fe] ratios. This scatter
is limited to the giant stars around solar metallicity. The giants also appear to be much more concentrated on the thin disk plane.
\item At low metallicity ([Fe/H]$<-0.5$) [Zn/Fe] appears constant at all galactocentic radii, and slightly supersolar.
\item As higher metallicities, [Zn/Fe] decreases to the solar value for stars roughly outside R$_{\rm GC}>7$kpc.
\item Conversely, stars at R$_{\rm GC}<7$kpc show an increasing depletion of Zn with increasing metallicity, 
down to about [Zn/Fe]=$-0.3$ for stars with [Fe/H]$>0$, despite with a significant dispersion. This behavior is in 
agreement with the low [Zn/Fe] values found in the Milky Way Bulge giants by \citet{barbuy15}. 
\item The low [Zn/Fe] observed in the (inner) thin disk giants can tentatively be explained  as due 
to dilution from almost Zn-free SN Ia ejecta, since a compatible level of dilution is observed in Ca and S. 
However, the observed [Mg/Fe] is 0.12 dex too high with respect to what our simple calculation indicates.
\end{itemize}


\begin{acknowledgements}
The authors are grateful to A. Recio Blanco and P. de Laverny for stimulating discussions regarding the evolution of Zinc in the solar neighborhood.
This work was based on data products from observations made with ESO Telescopes at the La Silla Paranal Observatory under programme ID 188.B-3002. 
These data products have been processed by the Cambridge Astronomy Survey Unit (CASU) at the Institute of Astronomy, 
University of Cambridge, and by the FLAMES/UVES reduction team at INAF/Osservatorio Astrofisico di Arcetri. 
These data have been obtained from the Gaia-ESO Survey Data Archive, prepared and hosted by the Wide Field Astronomy Unit, 
Institute for Astronomy, University of Edinburgh, which is funded by the UK Science and Technology Facilities Council.
This work was partly supported by the European Union FP7 programme through ERC grant number 320360 and by the 
Leverhulme Trust through grant RPG-2012-541. We acknowledge the support from INAF and Ministero dell' Istruzione, 
dell' Universit\`a e della Ricerca (MIUR) in the form of the grant "Premiale VLT 2012 and PRIN-INAF 
2014". The results presented here benefit from discussions held during the Gaia-ESO workshops and conferences 
supported by the ESF (European Science Foundation) through the GREAT Research Network Programme.
Support for S. D. and L. Sbordone was provided by the Chile's Ministry of Economy, Development, 
and Tourism's Millennium Science Initiative through grant IC120009, awarded to The Millennium Institute of Astrophysics, MAS.
The project was funded by FONDATION MERAC. 
S. S. is supported by the European Commission through a Marie 
Sklodowska-Curie Fellowship, project PRIMORDIAL, grant agreement: 700907.
S. M. A. and S. A. K. acknowledge the SCOPES grant No. IZ73Z0-152485 for financial support. L.M. acknowledges support from ``Proyecto interno" \#803 of the Universidad Andres Bello.
M. T. C. acknowledges the financial support from the Spanish Ministerio de Economía y Competitividad, through grant AYA2013-40611-P.
T. B. and S. F. were funded by the project grant ``The New Milky Way'' from Knut and Alice Wallenberg Foundation.
S. G. S. acknowledges the support by Funda\c c\~ ao para a Ci\^ encia e Tecnologia (FCT) through national funds and a research grant 
(project ref. UID/FIS/04434/2013, and PTDC/FIS-AST/7073/2014). S.G.S. also acknowledges the support from FCT through Investigador 
FCT contract of reference IF/00028/2014 and POPH/FSE (EC) by FEDER funding through the program Programa Operacional de 
Factores de Competitividade -- COMPETE.
L. Spina acknowledges the support from FAPESP (2014/15706-9).
R. S. acknowledges support the Polish Ministry of Science and Higher Education.
\end{acknowledgements}

\bibliographystyle{aa}

\end{document}